\documentclass[12pt]{article}
\usepackage[]{cite}
\usepackage[]{epsfig}

\newcommand{\hepnumber}{hep-ex/0101015}
\newcommand{\ee}{\mbox{${\mathrm{e}}^+ {\mathrm{e}}^-$}}
\newcommand{\tautau}{\mbox{$\tau^+\tau^-$}}
\newcommand{\mm}{\mbox{$\mu^+\mu^-$}}

\newcommand{\ra}        {\mbox{$\rightarrow$}}   

\newcommand{\qq}         {\mbox{$\mathrm{q}\bar{\mathrm{q}}$}}

\newcommand{\bb}         {\mbox{$\mathrm{b}\bar{\mathrm{b}}$}}
\newcommand{\cc}         {\mbox{$\mathrm{c}\bar{\mathrm{c}}$}}
\newcommand{\ff}         {\mbox{$\mathrm{f}\bar{\mathrm{f}}$}}
\newcommand{\nunu}       {\mbox{$\nu\bar{\nu}$}}

\newcommand{\mZ}         {\mbox{$m_{\mathrm{Z}}$}}
\newcommand{\mW}         {\mbox{$m_{\mathrm{W}}$}}
\newcommand{\mH}         {\mbox{$m_{\mathrm{H}}$}}
\newcommand{\mh}         {\mbox{$m_{\mathrm{h}}$}}
\newcommand{\mA}         {\mbox{$m_{\mathrm{A}}$}}
\newcommand{\chpone}{\tilde{\chi}^+_1}
\newcommand{\ntone}{\tilde{\chi}^0_1}
\newcommand{\nttwo}{\tilde{\chi}^0_2}
\newcommand{\chmone}{\tilde{\chi}^-_1}

\newcommand{\co}{\mbox{${\tilde{\chi}_1^0}$}}
\newcommand{\ct}{\mbox{${\tilde{\chi}_2^0}$}}

\newcommand {\Ho}        {\mbox{$\mathrm{H}^{0}$}}
\newcommand {\Ao}        {\mbox{$\mathrm{A}^{0}$}}
\newcommand {\ho}        {\mbox{$\mathrm{h}^{0}$}}
\newcommand {\Zo}        {\mbox{$\mathrm{Z}^{0}$}}

\newcommand{\qqp}{\mbox{$\mathrm{q\overline{q}^\prime}$}}

\newcommand{\tnt}{\mbox{${\tau\nu_{\tau}}$}}

\newcommand{\mHpm}{\mbox{$m_{\mathrm{H}^{\pm}}$}}

\newcommand{\lpair}{\mbox{$\ell^+\ell^-$}}

\newcommand {\Wpm}       {\mbox{$\mathrm{W}^{\pm}$}}

\newcommand{\gaga}       {\mbox{$\gamma\gamma$}}
\newcommand{\WW}         {\mbox{$\mathrm{W}^+\mathrm{W}^-$}}
\newcommand{\wen}        {\mbox{$\mathrm{W}^+\mathrm{e}^-\overline{\nu}_e$}}

\newcommand{\ipb}         {\mbox{pb$^{-1}$}}

\newcommand{\slsl } {\mbox{$\tilde\ell^+\tilde\ell^-$}}
\newcommand{\slp} {\mbox{$\tilde\ell^+$}}
\newcommand{\sqrts}     {\sqrt{s}}

\begin{document}
\begin{titlepage}
\vspace{-1cm}
\begin{flushright}
OPAL Conference Report CR460 \\
\today \\
\hepnumber
\end{flushright}
\bigskip\bigskip
\begin{center}
{\LARGE\bf Searches at LEP} \\
\bigskip\bigskip
{\Large Tom Junk} \\\bigskip
{\it Carleton University \\ 1125 Colonel By Drive, Ottawa, Canada K1S 5B6}\\ 
\bigskip\bigskip\bigskip
{\large Abstract}
\end{center}
Searches have been conducted for a broad range of new phenomena by the four
experiments ALEPH, DELPHI, L3, and OPAL, at LEP2.  Each experiment contributes
approximately 150~\ipb\ of \ee\ annihilation data with a mean $\sqrts$ of 205.9~GeV
in 2000 to these searches (data prepared for the September 5 LEPC meeting).
The statistical procedure for setting limits and
evaluating the significance of excesses observed in the data is reviewed.
Search results are presented for the Standard Model Higgs boson, the neutral
Higgs bosons in the MSSM, charged Higgs bosons, invisibly decaying Higgs bosons
produced by Higgs-strahlung, and fermiophobic Higgs bosons.  Search results
are briefly summarized for gauginos, stops, and staus.  The photon recoil spectrum
is checked for hints of new physics.  
\bigskip\bigskip\bigskip\bigskip
\begin{center}
{\large Results presented here have been prepared by the four LEP Collaborations, ALEPH, DELPHI, L3, and OPAL,
and by the LEP Higgs and SUSY working groups, for presentation at the September 5 open meeting of the
LEP experiments committee (LEPC).}
\end{center}
\bigskip
\begin{center}
{\large Presented at the 5$^{\mathrm{th}}$ International Symposium on Radiative~Corrections \\
Carmel, California, September 11--15, 2000}
\end{center}
\end{titlepage}

\section{Introduction}

In 1995, LEP finished taking large samples of data on the \Zo\ resonance
and began a program of increasing the beam energy in order to study in detail
known phenomena at higher energies and to search for new particles and
interactions.  The very highest beam energies were reached in 2000,
and in that year each experiment collected $\approx 150$~\ipb\ of \ee\ annihilation
data at an average $\sqrts$ of approximately 205.9~GeV
for inclusion in the results presented at the September~5, 2000 meeting of 
the LEP Experiments Committee (LEPC).  In addition, the searches for the SM and
MSSM Higgs bosons were combined by the LEP Higgs Working Group and shown
at the LEPC meeting and are summarized here.  The search for a neutral Higgs boson
produced in association with a \Zo\ boson is emphasized here because of recent
interest generated by excess events observed by the ALEPH collaboration in the
summer of 2000.  In addition, a brief selection of searches for supersymmetric
particles is given, and a summary of the running strategy for the rest of 2000
is presented.

The main Standard Model backgrounds
for the searches described here are
\begin{itemize}
\item Two-photon processes, \ee\ra\ee\ff, where the \ff\ pair has a very low
invariant mass and is produced in $t$-channel exchange of two photons.
\item Radiative returns to the \Zo.  Initial state 
radiation (ISR) reduces the effective $\sqrts$ to
the \Zo\ pole energy and produces boosted \Zo\ events.  The photon usually
escapes down the beampipe, but may be observed, and the probability of two
hard photons increases with increasing $\sqrts$.  These events are backgrounds
to analyses requiring missing energy, and electromagnetic coverage is needed very close
to the beam axis in order to suppress this background to searches requiring missing
transverse momentum.
\item W-pair production \ee\ra\WW.  These produce four-jet final states,
or jets+a lepton+missing energy, or two leptons and missing energy.  The
last is classified as an ``acoplanar dilepton'' because the plane containing the
two leptons does not in general contain the beam axis.  Very few b quarks are produced
in W decay and so this background can be suppressed in Higgs searches with b-tags with
good background rejection.  Searches for acoplanar dileptons have \WW\ production
as an irreducible background in some kinematic regions, while the two-photon background
dominates in other regions.
\item Z-pair production, \ee\ra\Zo\Zo.  These events may have four jets, two jets
and missing energy, two jets and two like-flavor but opposite-sign leptons, four leptons,
or two leptons and missing energy.  They have a high b quark content and constitute
a primary irreducible background to Higgs boson searches.
\item Single W production, \ee\ra\wen.  This can produce jets+missing energy (``acoplanar
dijets''), jets+a forward
electron, or a lepton+missing energy or a lepton+a forward electron+missing energy.
\item Single Z production \ee\ra\Zo\ee.  This process can produce acoplanar dijets or
acoplanar dileptons if the electrons are not detected.  One or both of the electrons
may be detected however.
\end{itemize}

\section{Statistical Procedure}

A single procedure is used by the four experiments and by the LEP
Higgs and SUSY working groups to determine whether a model
of new physics is excluded by LEP search data or if the background
hypothesis is disfavored relative to a particular signal hypothesis.  The procedure
begins with a full specification of the model to test, with the
masses, production cross-sections, and decay branching ratios
specified.  If the model under test allows many possibilities for
the parameters, then in general the parameters are scanned and
excluded regions are given in the parameter space that is allowed.
For a specific model with a specific choice of its free parameters,
histograms of the expected signal, background, and data events
are formed in variables that separate the expected signal from the
expected background.  Then all possible experimental outcomes
are considered, and they are ordered according to those that
are more signal-like (more candidate events in bins where a signal
is expected), and those that are more background-like
(fewer such candidates).  The variable used for ordering the
outcomes is the ``likelihood ratio," the ratio of the probability
of observing the particular outcome in the signal+background hypothesis
to the probability of observing the same outcome in the background-only
hypothesis:
\begin{equation}
Q=\frac{P_{poiss}(data|signal+background)}{P_{poiss}(data|background)}.
\end{equation}
Each bin of each histogram may be considered as an independent counting
experiment, and the Poisson probabilities multiply.  The expression 
for $\log Q$ is convenient for combinations of
many bins in many experiments:
\begin{equation}
\log Q=\sum_{i}\left(n_i^{data}\log\left(1+\frac{s_i}{b_i}\right)-s_i\right),
\end{equation}
where $s_i$ is the signal estimation for a bin of a search channel,
$b_i$ is the background estimation, $n_i^{data}$ is the number of observed
data events, and the sum runs over all bins in all search channels.
Some searches may have just one bin in them, while others may have some regions
of histograms of measured variables
with good separation of the signal from the background and other regions with
poorer separation.  Because $\log Q$ reduces to a sum of event weights,
events may be classified by the $s/b$ in the bins in which they appear.

A particular value of $\log Q$ will be obtained for each model hypothesis
and the available data and background estimates.  In order to determine
whether an outcome is sufficient to exclude that model, the probability
of obtaining that outcome in the signal+background hypothesis is computed:
\begin{equation}
{\mathrm{CL}}_{s+b} = P(Q\le Q_{obs}|signal+background).
\end{equation}
If ${\mathrm{CL}}_{s+b}<0.05$ then the signal+background hypothesis is ruled out
at the 95\% confidence level.  Another important confidence level to compute
is the consistency of the observation with the background hypothesis:
\begin{equation}
{\mathrm{CL}}_b=P(Q\le Q_{obs}|background),
\end{equation}
which is the probability of having observed no more than was observed, if
only background processes contribute.  This variable is used as the 
``discovery"
variable, requiring $1-{\mathrm{CL}}_b< 5.7\times 10^{-7}$ in order to claim a $5\sigma$
discovery.  These confidence levels may be computed using Monte Carlo
techniques~\cite{ref:clalex}, or by various convolution methods~\cite{ref:clfft},
\cite{ref:cltrj}.

One defect of the ${\mathrm{CL}}_{s+b}$ variable is that it is a test of the 
signal+background
hypothesis and not of just the signal hypothesis.  A consequence of this is that
a deficit of selected events relative to the background estimation
can rule out the background hypothesis
alone, and therefore also any signal+background hypothesis, even if the
signal is vanishingly small.  In fact, this is
expected to happen 5\% of the time at the 95\% confidence level.  In order
not to produce misleading limits or exclusions of parts of parameter space
to which the experiments are not sensitive, 
the following quantity is used~\cite{ref:clalex}:
\begin{equation}
{\mathrm{CL}}_s = {\mathrm{CL}}_{s+b}/{\mathrm{CL}}_b,
\end{equation}
which is expected to approach unity in the absence of sensitivity to a
particular signal.

\section{Standard Model Higgs Boson Searches}

The Standard Model Higgs boson is expected to be produced in \ee\ collisions
mainly by the Higgs-strahlung process when it is kinematically allowed,
and to a lesser extent by the \WW-fusion process.  The main attraction of
the latter process is that its cross-section does not drop rapidly near
$\sqrts-\mZ$, although its total rate is very small.  
The total production cross-section
near the kinematic edge is of the order of 50 to 500 fb,
depending on how close \mH\ is to $\sqrts-\mZ$.  The Standard Model
Higgs boson is expected to decay predominantly into \bb\ pairs in the mass
range of interest, with a branching ratio of 78\% at \mH=110~GeV and a branching
ratio of 74\% at \mH=115~GeV.
The second most important decay mode is to tau pairs, with a branching ratio of 
approximately
7\%, and the \WW\ decays take 8\%, which rises quickly with \mH.  Decays
to charm and gluons account for the remainder.  Efficient and pure b-tagging
is therefore important for search for Higgs bosons at LEP2.
The Standard Model Higgs search channels are differentiated by the 
\Zo\ decay mode that they select.  There is the four-jet channel 
(\Ho\ra\bb, \Zo\ra\qq), the missing-energy channel (\Ho\ra\bb, \Zo\ra\nunu),
the tau channels (\Ho\ra\bb,\Zo\ra\tautau\ and \Ho\ra\tautau,\Zo\ra\qq),
and the lepoton channels (\Ho\ra\bb, \Zo\ra\ee\ or \Zo\ra\mm).

Precision electroweak measurements may be used to estimate the value of the 
Higgs boson mass, assuming the Standard Model framework for radiative 
corrections.  The combined prediction of $\mH$ is $62^{+53}_{-39}$~GeV,
where the errors are symmetric and Gaussian in the variable 
$\log\mH$, as reported at ICHEP 2000~\cite{ref:EWICHEP}.  This prediction 
changes, however, when $\alpha_{EM}(\mZ)$ is computed differently or 
computed using additional low-energy \ee\ra\qq\ cross-section data from 
BES.

  In the absence of new particles and interactions,
the mass of the Standard Model Higgs boson can be from approximately
140~GeV to 180~GeV, where the lower bound arises from a vacuum stability
argument, and the upper bound from the requirement that the Higgs self-
coupling
remains finite at all energies.  For new physics interactions 
with a scale of the order of
a TeV, the Higgs boson mass is much less constrained by these arguments,
lying between 50 to 800 GeV~\cite{ref:pikpdg}.  Fine tuning 
arguments~\cite{ref:koldamurayama}, requiring that the magnitude of the 
radiative corrections to \mH\ are not too many orders of magnitude larger than 
\mH\ itself,
further restrict the possible ranges of \mH\ although these restrictions
are relaxed if new physics appears on the 1--10~TeV scale.

\begin{figure}[p]
\centerline{\epsfig{file=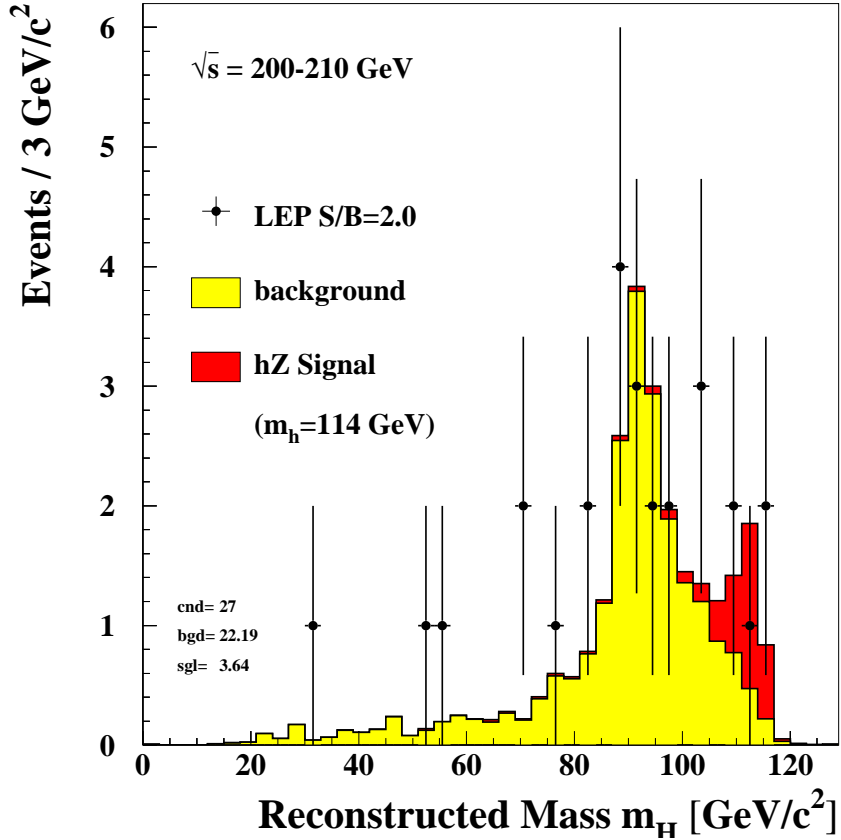,width=0.9\textwidth}}
\caption[]{\label{fig:mrectight}
Distribution of the reconstructed masses of candidates selected by the Higgs search
analyses summed over the four LEP experiments, summed over all search channels
for the data taken in 2000.  The selection cuts have been chosen such that
the integrated signal divided by the integrated background for reconstructed masses
above 109~GeV, for a SM signal hypothesis of 114~GeV, is roughly 2.0, in order to
keep the contributions from the four experiments roughly similar.  Each bin
contains contributions from several sources with different s/b.  The light histogram
shows the sum of all SM background expectations, the dark histogram shows the 
expected signal from a 114~GeV Higgs boson, and the points show the data.
}
\end{figure}

The combined distribution of the reconstructed masses in the four 
experiments'
Standard Model Higgs boson searches is shown in Figure~\ref{fig:mrectight} 
for a fairly tight set of selection requirements~\cite{ref:tullytalk}.  Along with the background 
expectation and the observed data counts is shown the expected signal from a 
Higgs boson
of mass 114~GeV.  The contributions to the histogram are given by experiment
in Table~\ref{tab:smexpt}.

The distribution of the reconstructed mass and the the total number of
selected events may be
uninformative or misleading in several ways.  The distribution of the
reconstructed mass is summed over different experiments, and over different search
channels at different center-of-mass energies.  The reconstructed mass 
resolutions
are different and depend on how close \mH\ is to the kinematic limit which
changes with $\sqrts$, and the relative amounts of the signal and background
are different from channel to channel, depending on the background
rates and signal branching ratios.  In a summed histogram of the reconstructed mass,
candidates in relatively clean channels are included in the same bins as
signal and background estimates from other channels with poorer performance.
Additional cuts have been applied after the standard analysis cuts
in order that the contributions from the 
different experiments are roughly equal in their total size.  If an experiment
has a large amount of the expected signal just failing the cut needed to make 
the reconstructed mass distribution plot, it may be more sensitive 
than another experiment with more expected signal on its side of its cut.  
Because all bins of all histograms in all variables (each search channel
from each experiment at each center of mass energy has its own histogram, 
and the variables can be the reconstructed mass, the b-tags, or combinations
of these and other information) can be combined using the uniform procedure 
outlined above, there is no loss of sensitivity in the
confidence level calculations, but an amount of information is necessarily lost 
when producing a histogram of the reconstructed mass and tables listing
its contents.

\begin{figure}[p]
\centerline{\epsfig{file=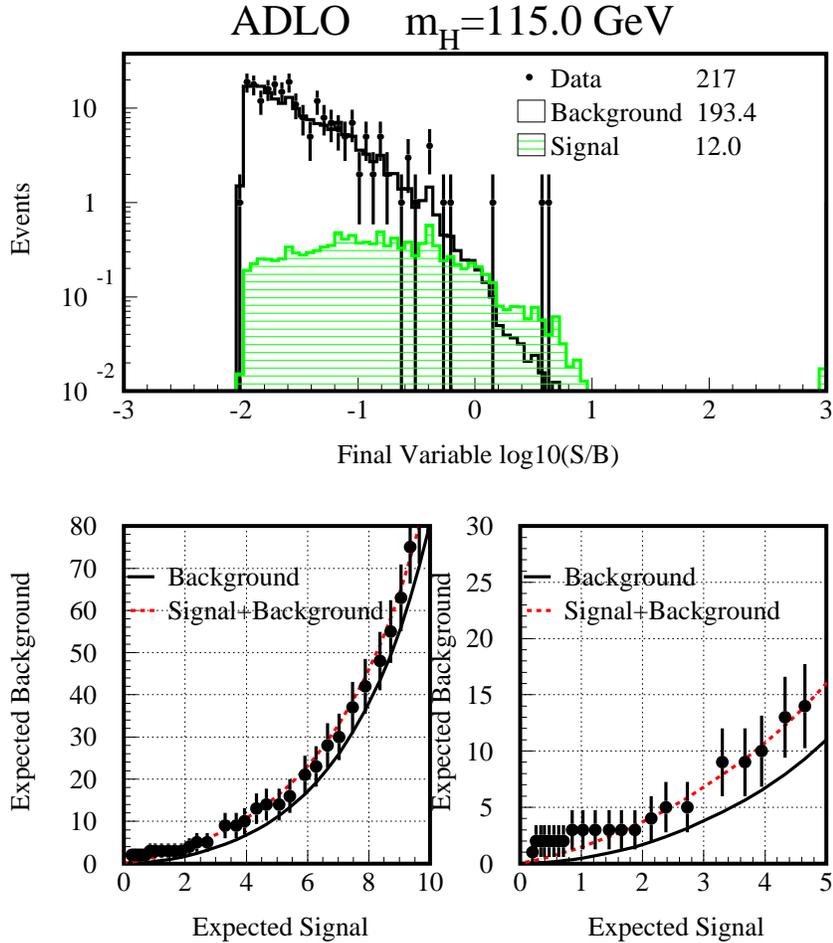,width=0.9\textwidth}}
\caption[]{\label{fig:sbplot}
Distribution of the $s/b$ for all bins of all search channels in all experiments
at all energies.  The signal is shown with a hatched histogram and the background
with the open histogram.  The data are shown with the points with error bars.
The most significant candidates from the ALEPH experiment's four-jet channels are the
three rightmost data points.  The lower graphs show the integral of the $s/b$ distribution
shown in the upper panel, from the high $s/b$ side.  The background integral is the solid curve,
the signal+background is the dashed curve, and the observed data are the points with error bars.
Neighboring points are highly correlated because of the cumulative sum.  The two lower panes show
the same integrals, but on different horizontal scales.
}
\end{figure}

The full amount of information in the searches is retained and displayed in a 
compact form if the histograms are rebinned in the variable $s/b$.  For each
bin of each histogram to be combined, the $s/b$ is uniquely determined.  
Because the test-statistic $\log Q$ is additive and depends only on the $s/b$
in the bins where the candidates are found and also on the total signal sum,
the contents of bins with the same $s/b$ may be simply added.  The result
is shown in Figure~\ref{fig:sbplot} along with its integral from the high
$s/b$ side~\cite{ref:tullytalk}.  The integral of this distribution at a particular cut in $s/b$
is the optimal answer to ``How many events are observed" and ``How many are
expected in the signal+background and the background-only hypotheses,"
for each possible setting of the cuts.

One observes in this distribution three events with rather high values of the 
local $s/b$.
These three are all four-jet candidates from ALEPH, with strong b-tags and
high reconstructed masses~\cite{ref:schlattertalk}. The selection of these 
events and the stability of their assigned significance has been checked in 
several ways.  A cut-based analysis is used to cross-check the primary neural-net-based
analysis and similar results are obtained.  All lower-energy data 
have been analyzed to look for biases in the reconstructed mass distribution 
towards a peak at the maximum kinematically allowed value, and no such bias is
seen.  The b-tag and neural net distributions also are modeled
well~\cite{ref:schlattertalk}.  One feature however, is that if all possible jet 
pairings are considered and events are removed if even one of these pairings 
yields jet-pair masses within 10 GeV of either \mW\ or \mZ, then the excess 
vanishes.  It was found in a Monte Carlo study, however, that the signal 
efficiency drops by 50\% by removing such events, and that this procedure does
not enhance the separation of signal from background.

There is a small excess observed in the DELPHI four-jet channel, but not in the
L3 or OPAL four-jet channels.  Also, no excesses are observed in the
combined missing-energy (\Zo\ra\nunu) channels, the lepton channels,
or the tau channels.  Combining all missing-energy, lepton, and tau channels 
together has about the same statistical power of all four-jet channels combined 
together.

\begin{figure}[p]
\centerline{\epsfig{file=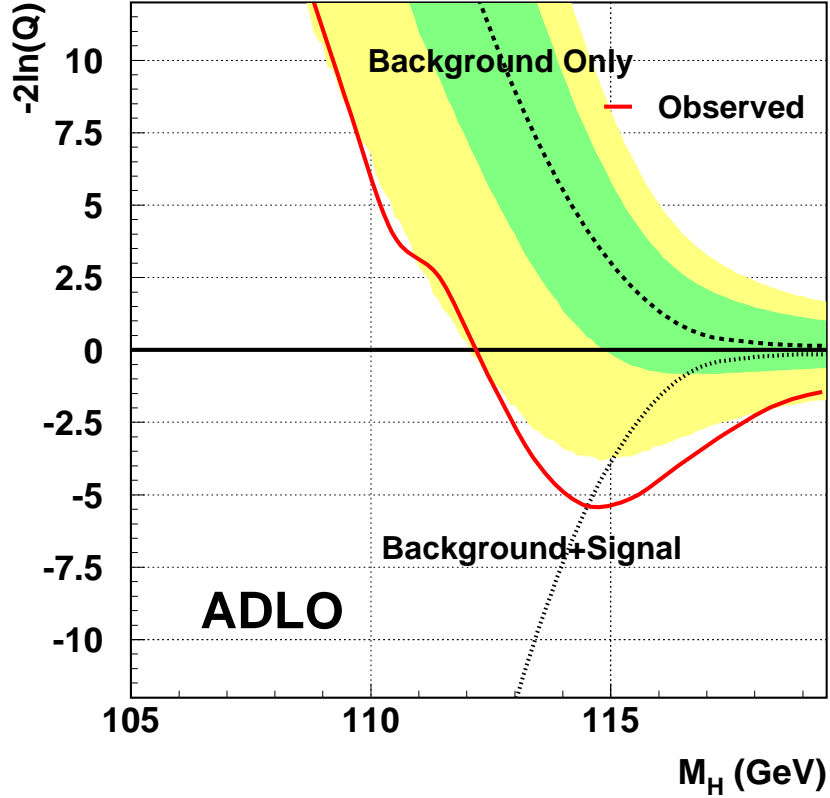,width=0.9\textwidth}}
\caption[]{\label{fig:sm2lnq} The value of the likelihood ratio
test statistic -2ln$Q$ as a function of the
test mass \mH.  Values below zero indicate a preference of the data for the signal hypothesis.  The solid
curve shows the observation in the combined LEP data, the dashed curve shows the median expectation in
an ensemble of background experiments, and the dotted curve shows the median expectation in an ensemble
of experiments in which the background and a signal originating from a SM Higgs boson of mass equal to
the test mass.  The dark band around the median background expectation is the 68\% probability interval
for the background ensemble, centered on the median expectation, and the light bands indicate the 95\% interval.
The minimum of the observed -2ln$Q$ curve is at \mH=115~GeV and has a value below zero, indicating that the
signal hypothesis is preferred.  The median
expectation from a 115~GeV Higgs boson is very close to the observed value.
}
\end{figure}

The combined test-statistic is shown in Figure~\ref{fig:sm2lnq}.  It has a 
minimum at \mH=114--115~GeV, and is the most compatible with the median 
expected signal at \mH=114~GeV.  The confidence levels are used to quantify 
how significant this observation is.

\begin{figure}[p]
\centerline{\epsfig{file=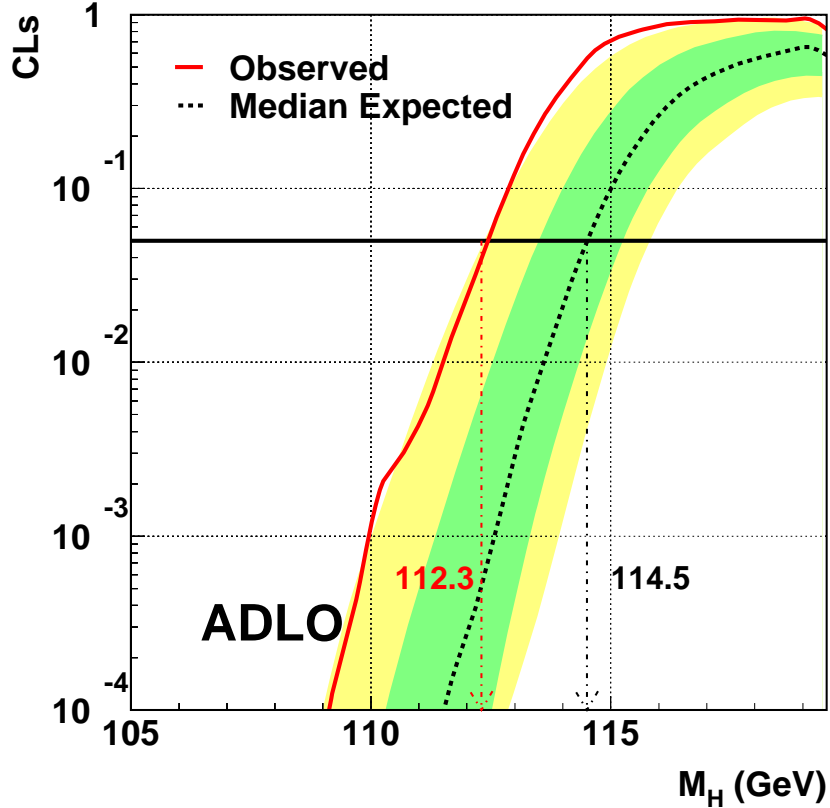,width=0.9\textwidth}}
\caption[]{\label{fig:smcls}  The exclusion confidence level ${\mathrm{CL}}_s$, as a function of the
test mass \mH.  The solid curve shows the value of ${\mathrm{CL}}_s$ computed from the selected events observed
in the data, for the four LEP experiments combined.  The dashed curve is the median expectation in
an ensemble of background-only experiments, and the dark and light shaded bands indicate the 68\% and 95\%
probability intervals around the expected median.  The 95\% CL exclusion limit is the lowest point
at which the observed ${\mathrm{CL}}_s$ crosses 0.05, and the median expected limit is where the median expectation
cross the line at 0.05.  A lower bound on \mH\ is obtained at 112.3~GeV, while the expected limit is 114.5~GeV.
}
\end{figure}

The exclusion limit is computed by finding the lowest \mH\ for which
${\mathrm{CL}}_s=0.05$.  It is computed for each experiment separately, and for 
each search channel separately, combining the results of the
experiments.  The expected limit is the median in a large 
ensemble of possible experiments in which only Standard Model background
processes contribute.  The value of ${\mathrm{CL}}_s$ and its expectation for the all 
channels combined from all experiments is shown in Figure~\ref{fig:smcls}. 
The exclusion limits are listed in Table~\ref{tab:smlimit}.
The branching ratios of the \Zo\ are well known and hence the separation by
channel is not that interesting, except for the tau channel, which covers
also Higgs boson decays to tau leptons.

\begin{figure}[p]
\centerline{\epsfig{file=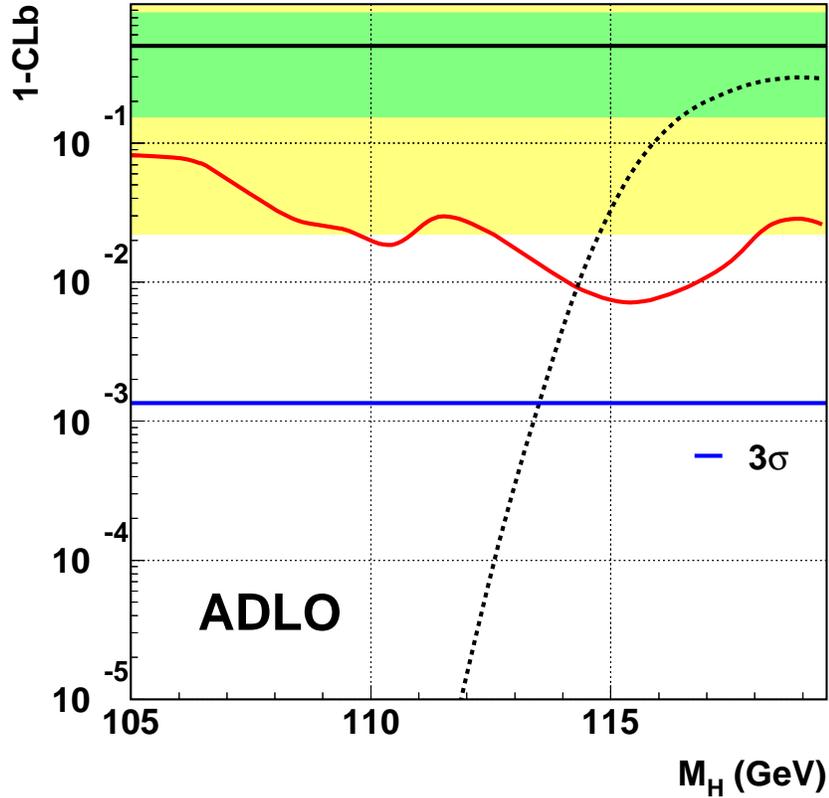,width=0.9\textwidth}}
\caption[]{\label{fig:smclb}  The background confidence level $1-{\mathrm{CL}}_b$ as a function of the test mass
\mH.  This is the probability of a fluctuation of the background to be at least as signal-like as observed at
that particular test mass; a small value indicates an excess of selected events.  If $1-{\mathrm{CL}}_b<5.7\times 10^{-7}$ then
a discovery may be claimed at the $5\sigma$ level.  The lowest $1-{\mathrm{CL}}_b$ observed in the data is
$7\times 10^{-3}$ at \mH$\approx$115~GeV, which corresponds to a significance of approximately $2.6\sigma$.
}
\end{figure}

In order to test for the compatibility of the observation with the expected 
background, $1-{\mathrm{CL}}_b$ is shown as a function of the tested \mH\ in 
Figure~\ref{fig:smclb}.  $1-{\mathrm{CL}}_b$ reaches its minimum at \mH=115~GeV, with 
a probability of consistency of the data with the background of 
$7\times 10^{-3}$, for a significance of approximately 2.6$\sigma$, due mainly
to the excess four-jet events in ALEPH, but also to events with lower values of 
$s/b$ which also contribute.

\begin{figure}[p]
\vspace{-1cm}
\centerline{\epsfig{file=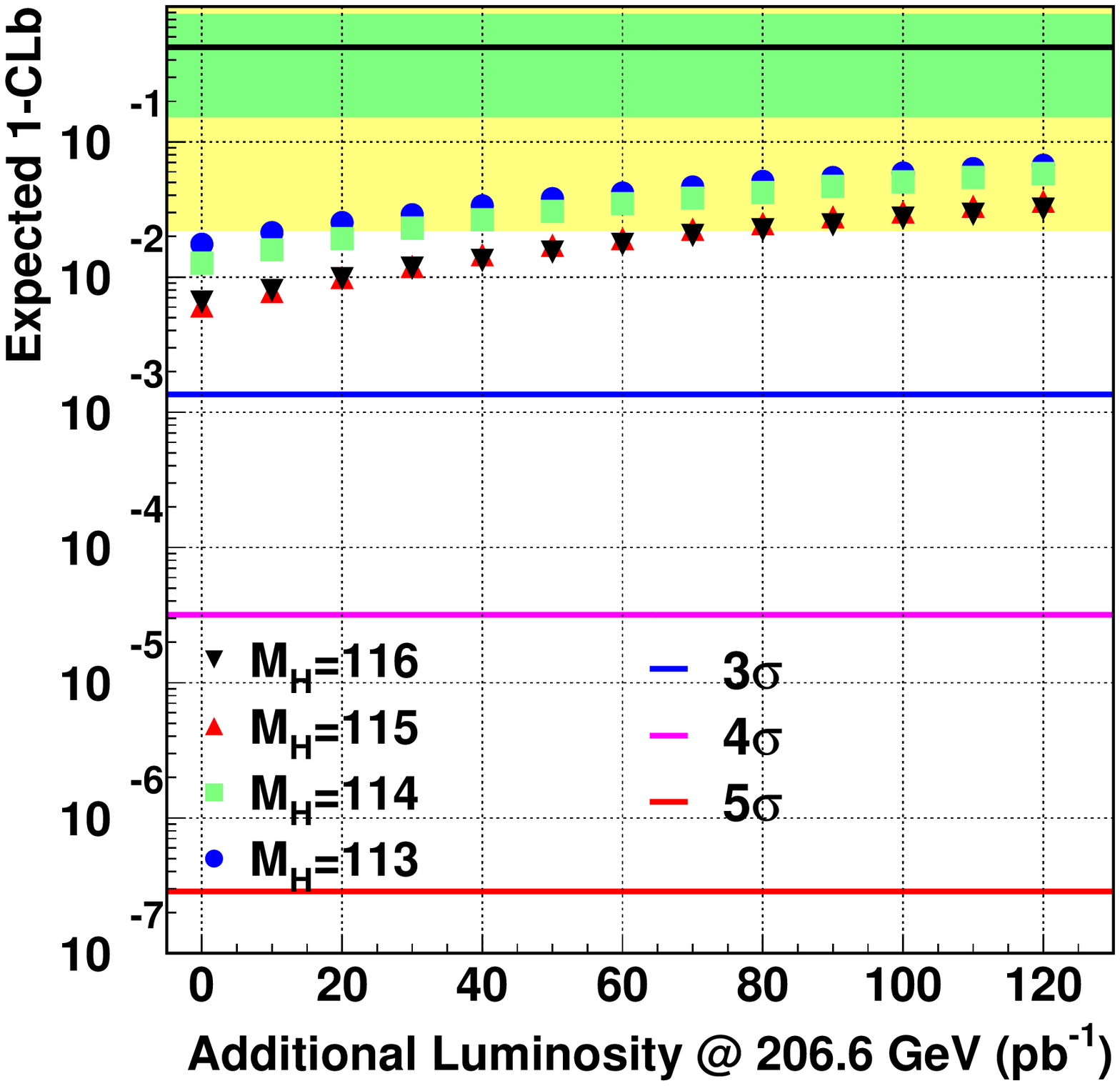,width=0.55\textwidth}}
\centerline{\epsfig{file=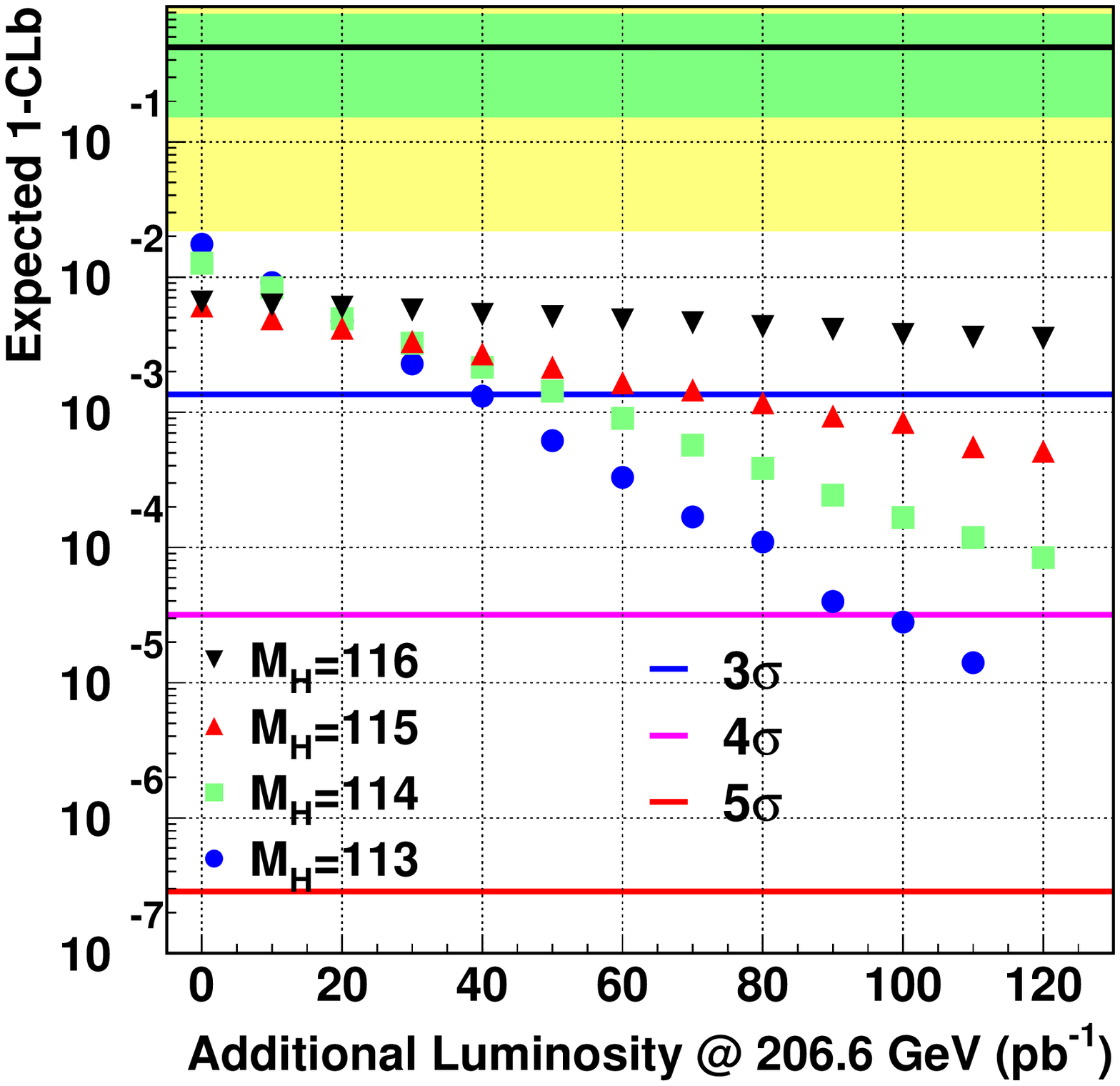,width=0.55\textwidth}}
\caption[]{\label{fig:smextension} Scenarios for additional LEP running in 2000 beyond the September 5 LEPC.
The top pane indicates the expected behavior of $1-{\mathrm{CL}}_b$ as a function of the amount of luminosity collected,
assuming a beam energy of 206.6~GeV and the absence of a signal, for four different values of the test mass.  This
quantifies the rapidity with which a background fluctuation should disappear with additional data accumulated.
The lower pane indicates the speed with which the significance of an excess will grow with time if the signal
were actually present, for different choices of the Higgs boson mass hypothesis \mH.  A 113~GeV Higgs boson
would be discoverable with a few months of extra running, but to extend the sensitivity out to 115~GeV requires
a run in 2001.  LEP typically collects in excess of 1~pb$^{-1}$ per day.
}
\end{figure}

In the case that there is no signal truly present, the excess would take 
approximately 60~\ipb\ per experiment at $\sqrts=206.6$~GeV
to fade away to a 2$\sigma$ excess for test mass
hypotheses near 115 GeV.  On the other hand, if the Higgs boson does
have Standard Model couplings and branching fractions, then one would
expect the significance of the excess to increase as more data are collected.  
As can be seen in Figure~\ref{fig:smextension}, it would take approximately
100~\ipb\ of data at $\sqrts=206.6$~GeV to obtain a 4$\sigma$ effect
for \mH=113~GeV, and around 140~\ipb\ for \mH=114~GeV.  On the other hand, 
significances of 3$\sigma$ can be obtained within 60~\ipb\ for \mH\ all the way
up to 115~GeV, and that amount can be collected in approximately 60~days
of running.

\section{Searches for Non-Standard Higgs Bosons}

\subsection{Neutral Higgs Bosons in the MSSM}

One of the simplest extensions to the Higgs sector of the Standard Model is to 
add a second Higgs field doublet.  The Minimal Supersymmetric Extension of the SM
(MSSM) requires this structure.  One field couples to up-type quarks and the 
other to down-type quarks, and there is a mixing angle $\alpha$ between these 
two fields in order to produce the physical Higgs states, which number five:
the \ho, the \Ao, the \Ho, and two charged Higgs bosons H$^+$ and H$^-$.  The 
ratio of the vacuum expectation values of the two fields is denoted $\tan\beta$.  
In the CP-conserving, low-energy effective MSSM studied 
here~\cite{benchmarks}, the remaining 
parameters are the mass of the \Ao, the CP-odd Higgs boson (the other two 
neutral bosons are CP even), the mass scale of the sfermions 
$M_{\mathrm{SUSY}}$ (here set to 1~TeV), the Higgs mass matrix parameter 
$\mu$ (here set to -200 GeV), the gaugino mass parameter $M_2$ (here set 
to 200 GeV), and the amount of stop mixing, chosen here to be zero or 
maximal.  Recent calculations of \mh\ including the dominant 2-loop terms are 
used~\cite{weigheiholl}.  The gluino mass is also a free parameter; it affects 
the Higgs masses and branching ratios through radiative corrections.

For the case of maximal stop mixing, and the choices of the other parameters 
given above, the value of \mh\ assumes its maximal value\footnote{Another 
calculation~\cite{carenawagner} using a renormalization-group improved one-
loop calculation, does in fact give slightly higher values of \mh\ for 
$\tan\beta<1$, but also smaller values of \mh\ for $\tan\beta>1$ -- we choose 
the calculation which gives the lowest upper end of the excluded $\tan\beta$ 
region.} as a function of $\tan\beta$ and is used to set conservative limits on 
$\tan\beta$.  This scenario is called the \mh-max scenario.

\begin{figure}[p]
\vspace{-0.5cm}
\centerline{\epsfig{file=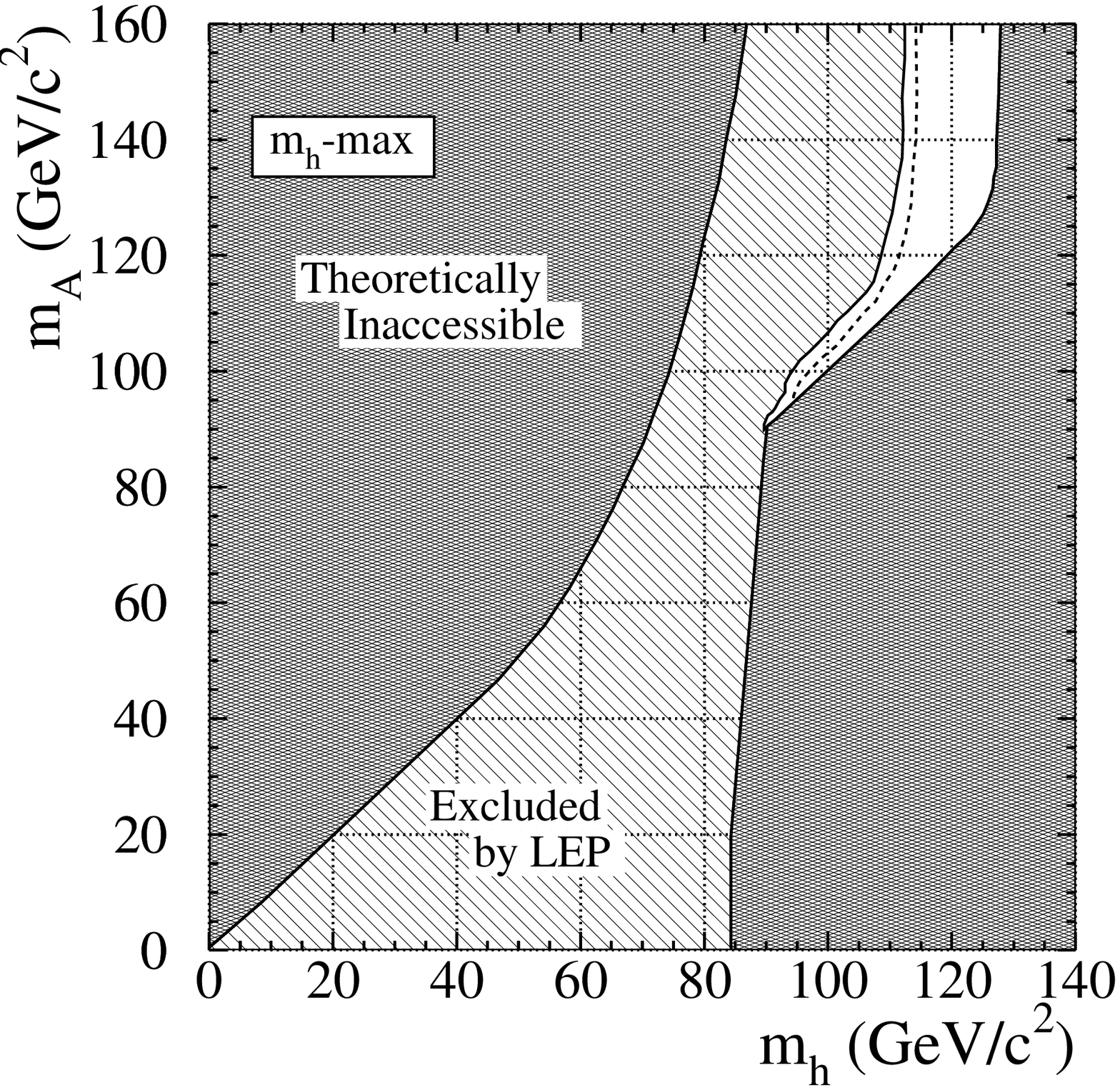,width=0.4\textwidth}\epsfig{file=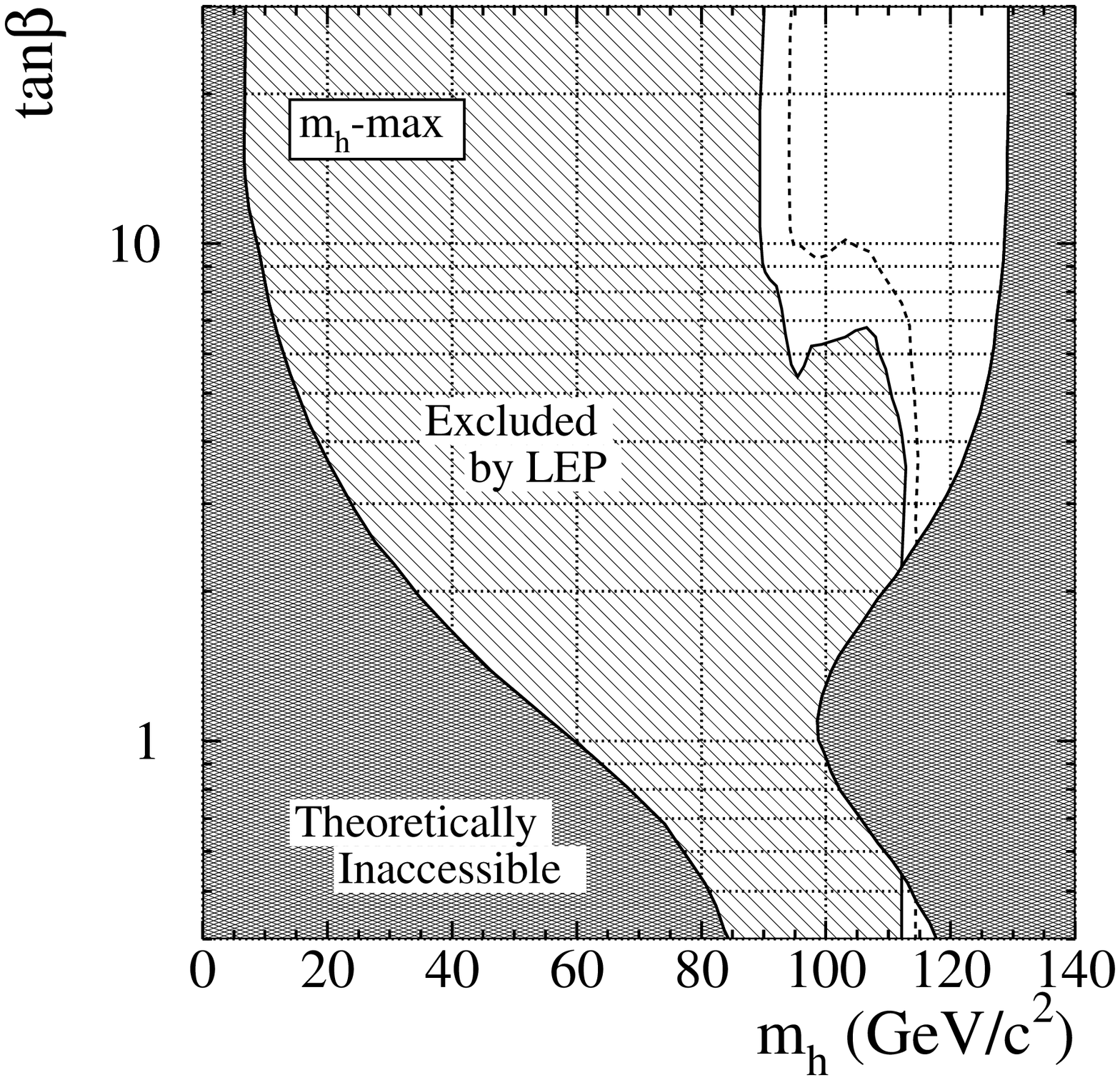,width=0.4\textwidth}}
\centerline{\epsfig{file=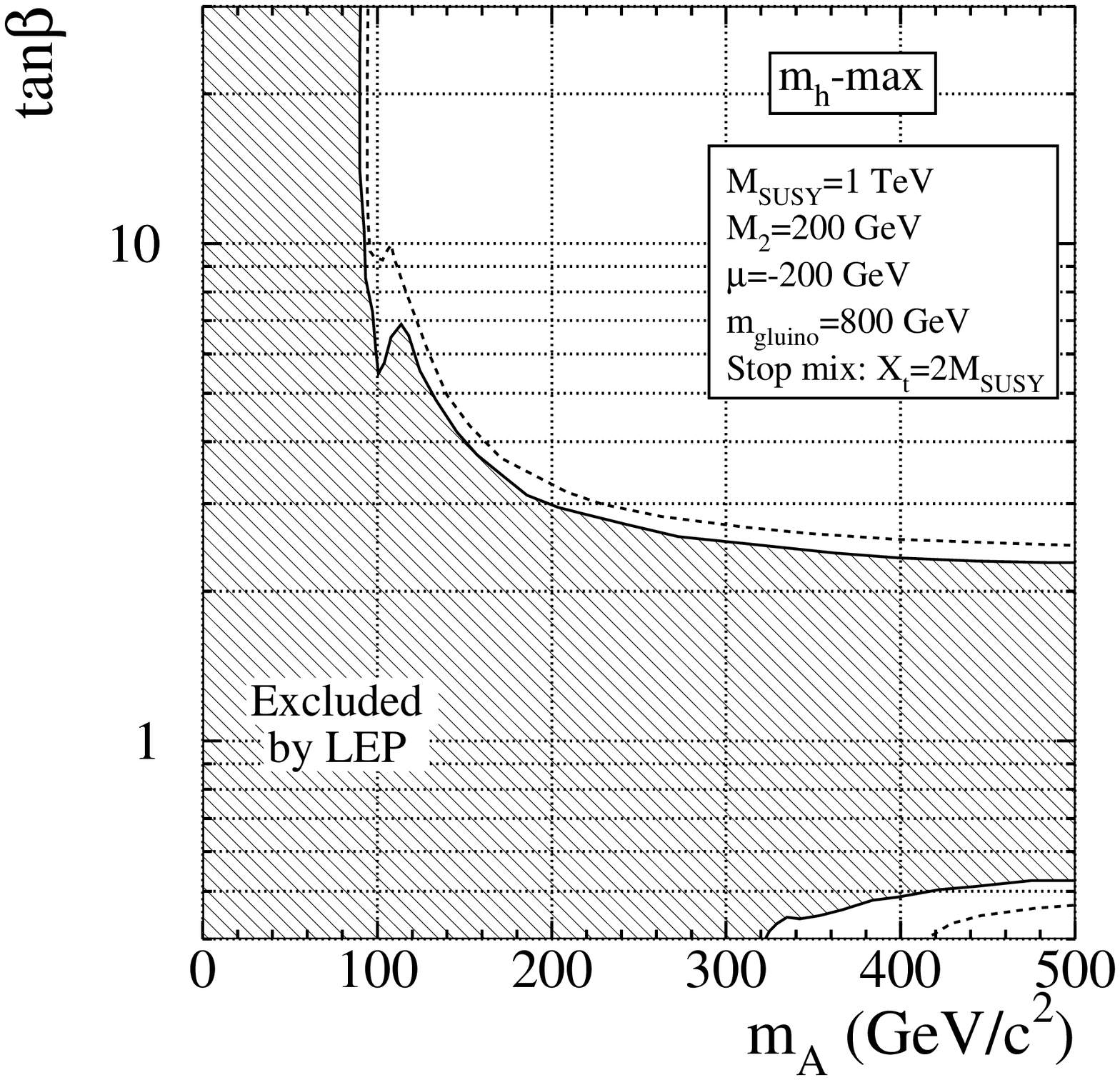,width=0.7\textwidth}}
\caption[]{\label{fig:maxmix}
Limits on \mh, \mA, and $\tan\beta$ in the \mh-max MSSM benchmark
scenario, described in the text.  The limits are shown in the (\mh, \mA) plane
(upper left), the (\mh, $\tan\beta$) plane (upper right), and in the (\mA, $\tan\beta$) plane (lower plot).
The excluded regions are shown with diagonal hatching, and the regions which are not allowed by the theory
are shown with dark hatchings.  The median expected boundaries of the excluded regions are shown with
dashed lines.
  The \mh-max scenario
is designed to give the most conservative limits on $\tan\beta$ from the intersection of the limit curve with the
theoretically unallowed region on the right of the (\mh, $\tan\beta$) plot.
}
\end{figure}

The searches used to set limits in this space are the same searches for the 
\ho\Zo\ final state used in the Standard Model section, but in addition, searches 
for \ho\Ao\ are performed in the \bb\bb\ and \bb\tautau\ final states.  The 
production cross-section for \ee\ra\ho\Ao\ is proportional to $\cos^2(\beta-
\alpha)$, which is largest for $\mh\approx\mA$ while the cross-section for 
\ee\ra\ho\Zo\ is proportional to $\sin^2(\beta-\alpha)$.  The cross-section
for \ho\Ao\ production grows more slowly along the diagonal $\mh=\mA$ than 
the \ho\Zo\ cross-section does for large \mA.  Therefore, the absolute lower 
limits on \mh\ will come from the case in which \ho\Zo\ production is 
suppressed, and the limits for \mA\ra$\infty$ are those obtained in the 
Standard Model, as can be seen from Figure~\ref{fig:maxmix}.

\begin{figure}[p]
\vspace{-1cm}
\centerline{\epsfig{file=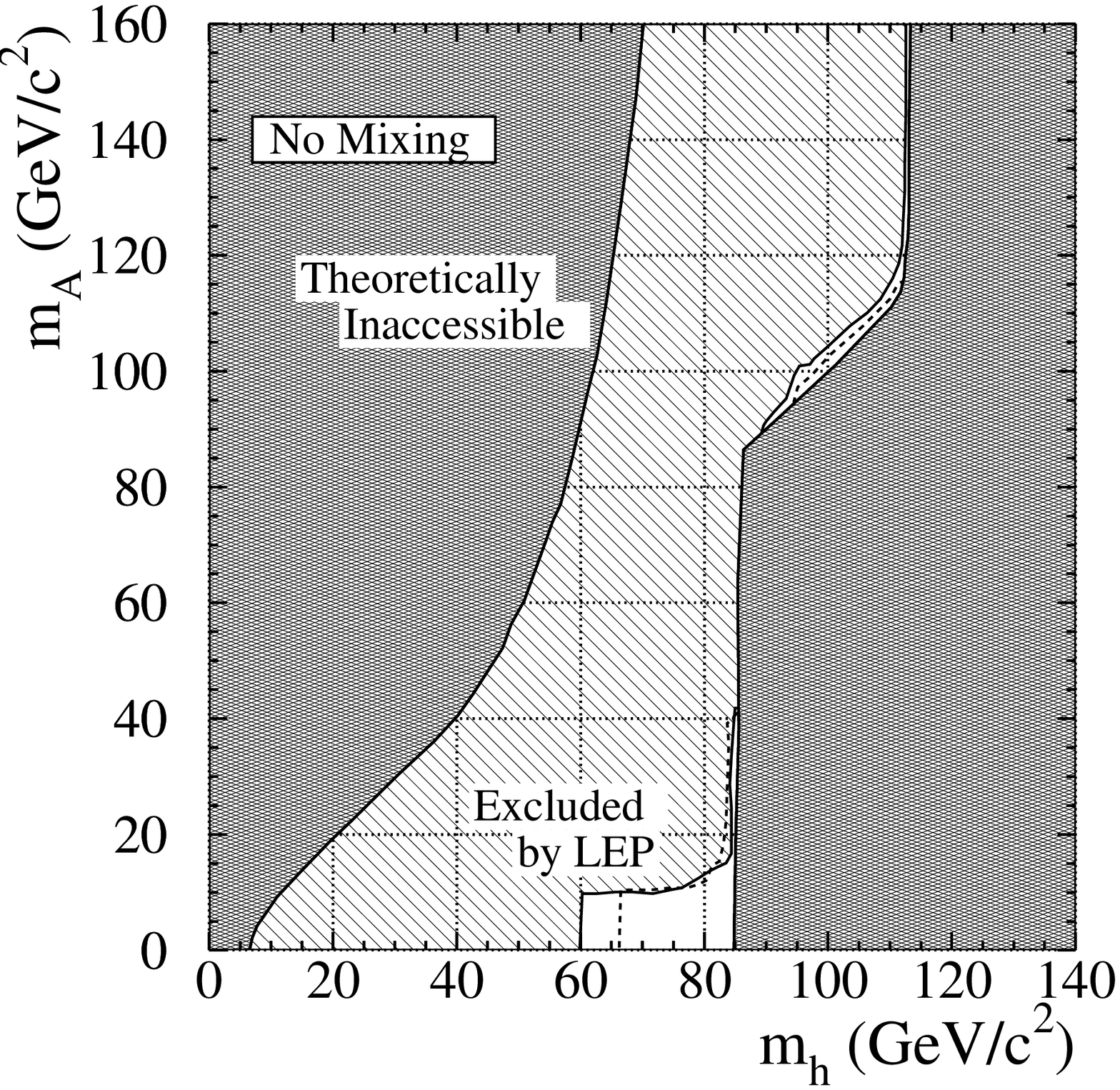,width=0.4\textwidth}\epsfig{file=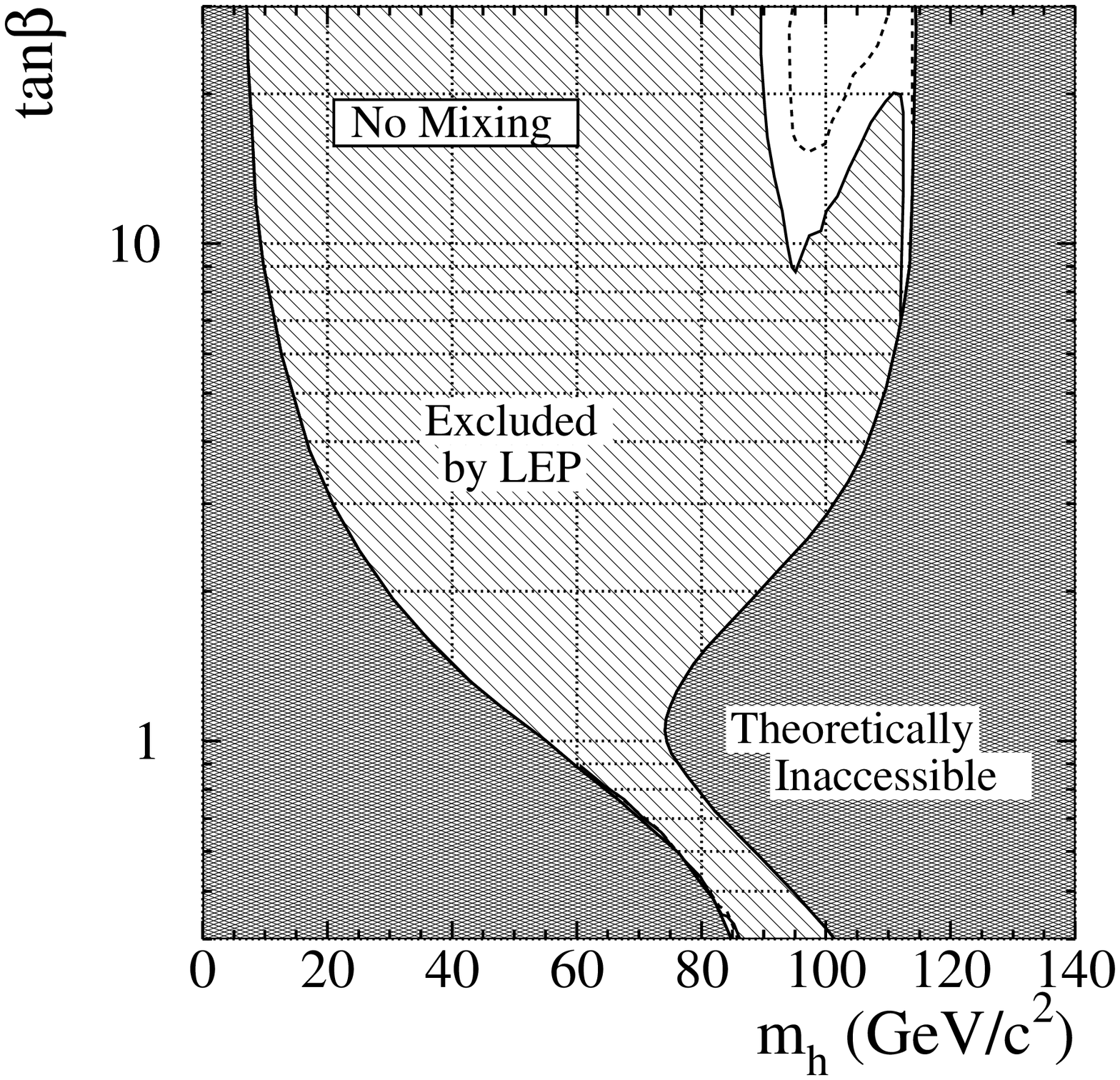,width=0.4\textwidth}}
\centerline{\epsfig{file=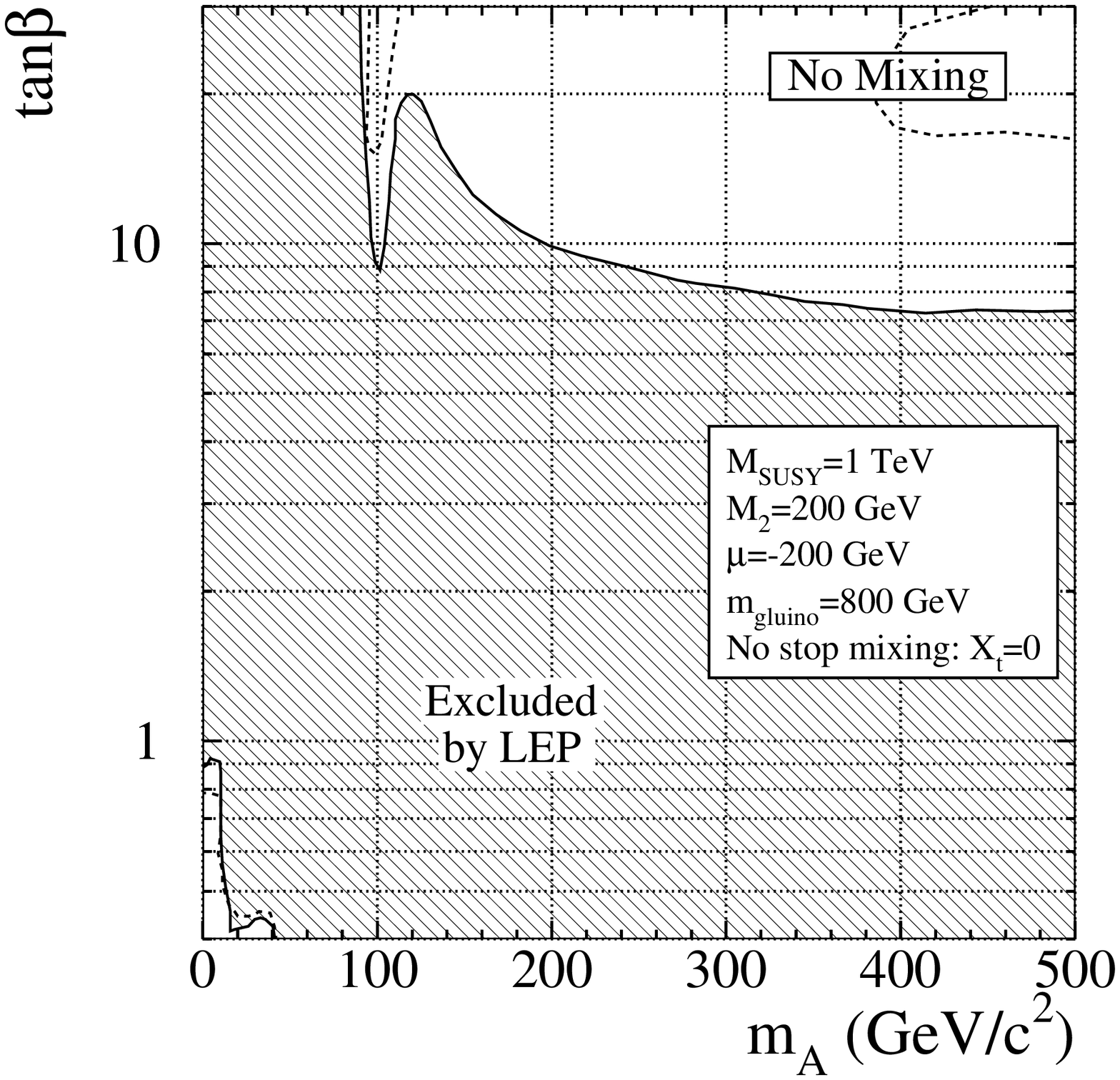,width=0.5\textwidth}}
\caption[]{\label{fig:nomix}  Exclusions in the no-stop-mixing MSSM benchmark scenario, described in the text.
The limits are shown in the (\mh, \mA) plane
(upper left), the (\mh, $\tan\beta$) plane (upper right), and in the (\mA, $\tan\beta$) plane (lower plot). 
The excluded regions are shown with diagonal hatching, and the regions which are not allowed by the theory
are shown with dark hatchings.  The median expected boundaries of the excluded regions are shown with
dashed lines.
  In this
scenario, the limits are similar for \mH$\approx$\mA and also the same for \mH\ at low $\tan\beta$ as they are in the
\mh-max scenario, but the limits on $\tan\beta$ from the intersection on the right-hand side of the
(\mh, $\tan\beta$) plot are much more stringent.  On the other hand, more parameter space is opened up at low
$\tan\beta$ for values of \mh\ between 60 and 85~GeV.  In this region, the \ho\ decays into \Ao\Ao\  and/or charm
and gluons, because the \bb\ decay is suppressed by the low value of $\tan\beta$, and the \bb\ decays of
the \Ao\ are also suppressed.  In this region, flavor-independent
searches, under development, will be used to search for possible signals, and if none are found, to exclude
the remaining part.
}
\end{figure}

For the case of no stop mixing, the maximal value of \mh\ as a function of 
$\tan\beta$ is much less, although a second problem opens up at low 
$\tan\beta$: the branching ratio for \ho\ra\bb\ can be suppressed by a larger 
decay width for \ho\ra\Ao\Ao, and for low $\tan\beta$ or low \mA, the decay rate
of the \Ao\ to \bb\ pairs is suppressed either by the coupling strength or the 
kinematics if $\mA<10$~GeV.  In this case, an "L"-shaped unexcluded region
opens up for low $\tan\beta$ and low $\mA$, shown in Figure~\ref{fig:nomix}.  
Additional flavor-independent searches, and searches specifically targeted at this 
region are being developed and will soon be included in the combination.  The 
limits on \mh\ and \mA\ presented in Table~\ref{tab:mhmalimits} ignore this 
unexcluded region, although the limits on $\tan\beta$ in the no-stop-mixing 
scenario include its effects.  In both the \mh-max scenario and the no stop 
mixing scenario, $\tan\beta$ is considered only up to 30, because for larger
values of $\tan\beta$ 
the \ho\ decay width can exceed the experimental mass resolution, and 
additional Monte Carlo signal samples are needed to assess the effect of 
lower $s/b$ in these channels.

\begin{figure}[p]
\centerline{\epsfig{file=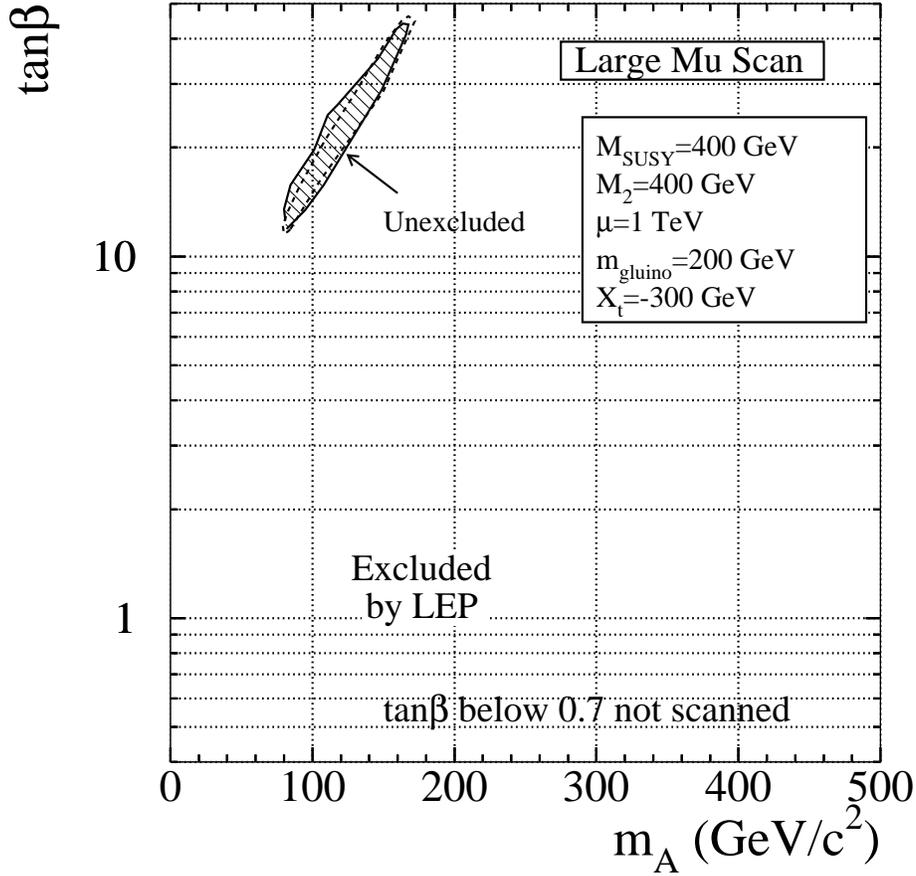,width=0.95\textwidth}}
\caption[]{\label{fig:largemu} The exclusion in the large-$\mu$ MSSM benchmark scenario, described in
the text.  Only the (\mA, $\tan\beta$) projection is shown because the unexcluded region is not easily visible 
in the other projections.  This scenario is designed to highlight portions of SUSY parameter space where the
\ho\ra\bb\ decay is suppressed, and the tau decays are not enhanced, the remainder being taken up by
W$^{(*)+}$W$^{(*)-}$ and \cc\ decays.  The unexcluded region is shown with diagonal hatching in this case, and
the median expected boundary of this region is shown with a dashed line.  For some points in this parameter space, the
heavy Higgs boson \Ho\ is kinematically accessible, and the searches for \ho\Zo\ are re-interpreted as searches for
\Ho\Zo\ where the latter searches have a better expected sensitivity.
}
\end{figure}

A third scenario has been proposed, called the ``large-$\mu$" 
scenario~\cite{benchmarks}, in which $M_{\mathrm{SUSY}}$ is taken to be 
400~GeV, $M_2=400$~GeV, $\mu=1$~TeV, and the gluino 
mass is 200~GeV.  This setting of parameters is designed to 
highlight loop effects which can suppress the decay \ho\ra\bb, without a 
corresponding enhancement of \ho\ra\tautau.  In this case, the \ho\ decays 
rather into gluons, charmed quarks, or W pairs, but only for high $\tan\beta$.
The decay widths of the \ho\ and the \Ao\ remain much smaller than the experimental
mass resolution up to $\tan\beta=50$.
The maximum value of \mh\ in this scenario is approximately 108~GeV, and for 
the region where $\sin^2(\beta-\alpha)$ is low and $\mh+\mA$ is kinematically 
out of reach at LEP2, the process \ee\ra\Ho\Zo\ is accessible with a
cross-section proportional to $\cos^2(\beta-\alpha)$.
Nearly all model points except those with difficult decay branching fractions
can therefore be excluded.
These difficult regions are at $\tan\beta>10$ and $80<\mA<180$, as seen in 
Figure~\ref{fig:largemu}.

\subsection{Charged Higgs Bosons}

\begin{figure}[p]
\vspace{-1cm}
\centerline{\epsfig{file=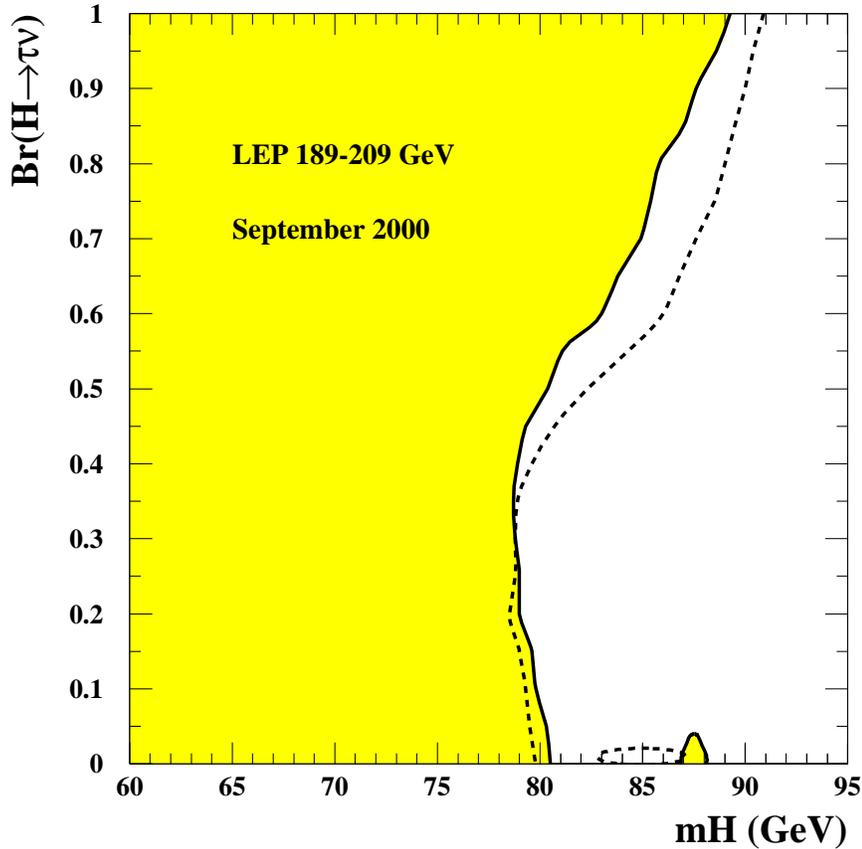,width=0.95\textwidth}}
\caption[]{\label{fig:chargedlimits}  Limits on the production of charged Higgs bosons, as a function of
the branching ratio Br(H$^+\ra\tau^+\nu_\tau$), assuming that
Br(H$^+\ra\tau^+\nu_\tau$)+Br(H$^+\ra$\qqp)=1.  The excluded region is shown with light shading and the boundary
is indicated with the heavy solid lines.  The boundary of the region expected to be excluded in 50\% of
background-only experiments is indicated by the dashed lines.  The background from \WW\ decays is more severe
for the four-jet search because of the \Wpm\ branching ratios, and impedes efforts to search for charged Higgs
bosons with masses close to the mass of the \Wpm.  As data accumulate, though, the sensitivity increase for
\mHpm\ significantly in excess of \mW, and a small island of exclusion is appearing in both the observed and
expected limits above \mW\ for Br(H$^+\ra\tau^+\nu_\tau$)=0.  The result is from the combination of the charged
Higgs boson searches from the four LEP experiments.
}
\end{figure}

At LEP, charged Higgs bosons are expected to be pair-produced via the
$s$-channel exchange of a $\gamma$ or a \Zo, and the production cross-section 
depends only on the mass of the H$^\pm$ and on well-measured electroweak 
parameters.  The decay modes of the H$^\pm$ are considered for the purpose of 
these searches to be limited to \qqp\ and \tnt.  The mass limits are produced 
therefore as a function of Br(H$^+\ra\tau^+\nu_\tau$).  The search channels 
include a four-jet search without b-tagging, a semileptonic search, and a fully 
leptonic search, in which the final state consists of an acoplanar pair of taus.
The predominant background is \ee\ra\WW, which can produce all of the 
available final states, although the acoplanar tau pair rate is significantly lower 
than the four-jet rate due to the branching ratios of the W.  The large \WW\ 
background sets the scale for the limits in the hadronic and semileptonic 
searches. The limits are shown in Figure~\ref{fig:chargedlimits}.  For 
Br(H$^+\ra\tau^+\nu_\tau$)=0, the observed
mass limit is 80.5~GeV and the median expected limit is 79.8~GeV.  For
Br(H$^+\ra\tau^+\nu_\tau$)=1, the observed
mass limit is 89.2~GeV and the median expected limit is 90.9~GeV.  The lowest limit 
obtained at any branching ratio is 78.7~GeV, with a median expectation of 78.5~GeV.
A small excluded ``island" appears in Figure~\ref{fig:chargedlimits} for
Br(H$^+\ra\tau^+\nu_\tau$)=0, where the search sensitivity above the \WW\ 
background peak is beginning to become sufficient to exclude a small 
region.  More data would allow this island to grow and eventually connect with the 
main excluded region, leaving a hole near \mW\ which can be excluded only 
with a larger amount of integrated luminosity.

\subsection{Searches for \Ho\ra$\gamma\gamma$}

\begin{figure}[p]
\centerline{\epsfig{file=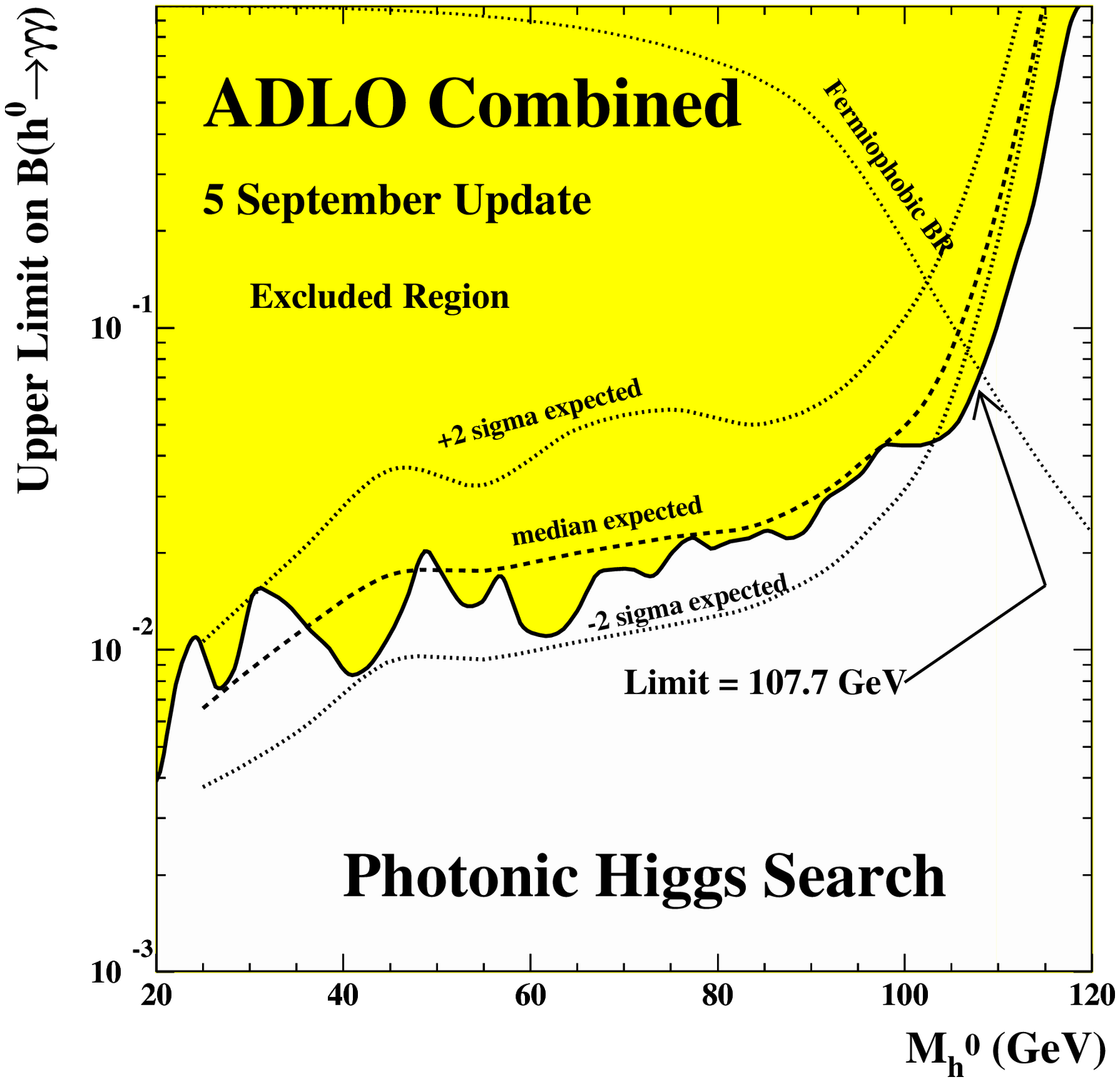,width=0.95\textwidth}}
\caption[]{\label{fig:gammagammalimits} 
Limits obtained on the production of
\ee\ra\Ho\Zo\ra\gaga\Zo, combining all search channels from the four LEP experiments
for all decay modes of the \Zo.  These searches are not combined with the searches for the non-photonic
decays of the Higgs boson, and are therefore limits on the production cross-section of a Higgs boson
which decays only into photons, relative to the Standard Model production cross-section.  Alternatively,
they are interpreted as limits on the photonic branching ratio of the Higgs boson, ignoring the other
search results.  A prediction of the photonic branching ratio of the Higgs boson in a model in which
the couplings of the Higgs boson to fermions are all zero is shown with the
dashed line.  In such a fermiophobic model, a lower mass limit of 107.7~GeV is obtained.
}
\end{figure}

The final states \qq\gaga, \lpair\gaga, and \gaga+missing energy are sought by 
the four LEP experiments and combined.  Because the branching ratio 
Br(\Ho\ra\gaga) is small in the Standard Model (of the order 10$^{-3}$),
mass limits cannot be set on the SM Higgs from this search only.
  This search is more interesting when considering models in 
which the \Ho\ decays are non-standard.  In particular, if the \Ho\ fails to couple 
to fermions entirely, then the available decay modes are into \gaga\ and \WW, 
the first of which proceeds only at the one-loop level mediated by a W boson.
As the mass of the \Ho\ increases, the \WW\ branching fraction gradually 
becomes more important and the mass limits obtained at LEP2 are mainly 
determined by this behavior than by the power of the searches.  The limits are 
expressed in Figure~\ref{fig:gammagammalimits} by assuming the SM 
production cross-section for \ee\ra\Ho\Zo, and by ignoring the results of searches
for other decays of the 
\Ho\ to set limits on the branching ratio Br(\Ho\ra\gaga).  Alternatively, this can 
be interpreted as a limit on the production cross-section as a fraction of the SM 
cross-section, with Br(\Ho\ra\gaga)=1.  In the fermiophobic model, the 
observed mass limit is 107.7~GeV, with a median expected limit of 105.8~GeV.

\subsection{\ho\ra\ Invisible Particles}

The Higgs boson may decay invisibly in the MSSM if the lightest 
neutralino has a mass of less than half the mass of the Higgs.
Two important advantages of an \ee\ collider are that the center-of-mass 
energy of each interaction is known with a high degree of precision, and that the 
total momentum is zero.  These features can be used to search for the process 
\ee\ra\ho\Zo, where the \ho\ decays invisibly, because the \Zo\ decay products 
can be measured and the missing mass can be inferred.  In this case, \Zo\ decays to 
quarks are exploited for the search.  Neutrino decays are not useful, and the 
leptonic decays have a low relative branching ratio.  Tau decays in particular 
pollute the leptonic sample because
the neutrinos in such events carry a large amount of missing energy.  

\begin{figure}[p]
\centerline{\epsfig{file=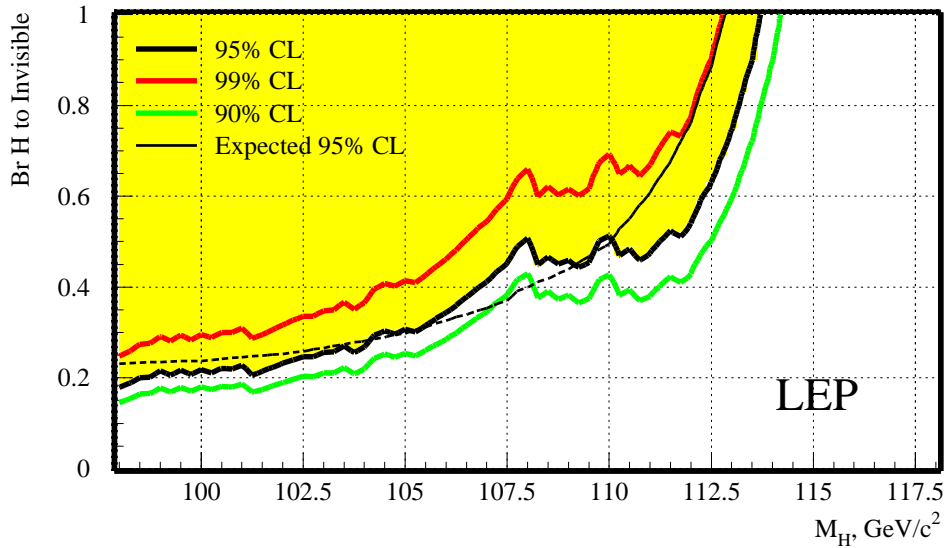,width=0.95\textwidth}}
\caption[]{\label{fig:invislimits} 
Limits obtained on the production of
\ee\ra\Ho\Zo, where the \Ho\ decays invisibly, combining all search channels from the four LEP experiments
for the included decay modes of the \Zo.  These searches are not combined with the searches for the visible
decays of the Higgs boson, and are therefore limits on the production cross-section of a Higgs boson
which decays only invisibly, relative to the Standard Model production cross-section.  Alternatively,
they are interpreted as limits on the invisible branching ratio of the Higgs boson, ignoring the other
search results.  A lower bound on a Higgs boson produced with the SM production cross-section and decaying
invisibly is set at 113.7~GeV, and the median expectation is 112.8~GeV.
}
\end{figure}

The limits are shown in Figure~\ref{fig:invislimits} assuming the Standard 
Model production cross-section and are limits on the invisible branching ratio 
of the Higgs, ignoring the results of searches for visible Higgs decays.  
Alternatively, these limits can be interpreted as limits on the production 
cross-section divided by the Standard Model production cross-section, assuming 
100\% invisible decays of the \ho.  For the SM cross-section and 100\% 
invisible decays, the mass limits obtained are 113.7~GeV (observed) and
112.8~GeV (median expectation).

\section{Gaugino, Squark, Slepton Searches}

Charginos may be produced either in the $s$-channel via photon or \Zo\ exchange,
or in the $t$-channel via exchange of an electron sneutrino.  These diagrams interfere
destructively, although the $t$-channel diagram is important only for light
electron sneutrinos.  The chargino may decay into a W and a neutralino, or
into a slepton and a neutrino, where the slepton decays into a neutralino and a lepton,
or directly into a lepton and a sneutrino.  All of these decay modes produce similar
final states -- two leptons (or jets) and missing energy.
The branching ratios for these processes
depend on the slepton and sneutrino masses, and the mass difference $\Delta M$
between the chargino and the LSP (either the neutralino or sneutrino) strongly affects the
final state kinematic distributions.  For a small mass difference, the visible decay
products of the chargino have low visible energies.  These final states are similar to 
the two-photon background processes which have large cross-sections in \ee\ collisions at high
energies.  For very large $\Delta M$, the final states resemble W$^+$W$^-$ production.
The search analyses are therefore optimized in separate regions of $\Delta M$ due to
the different makeup of the signal and background estimations.

\begin{figure}[p]
\vspace{-1cm}
\centerline{\epsfig{file=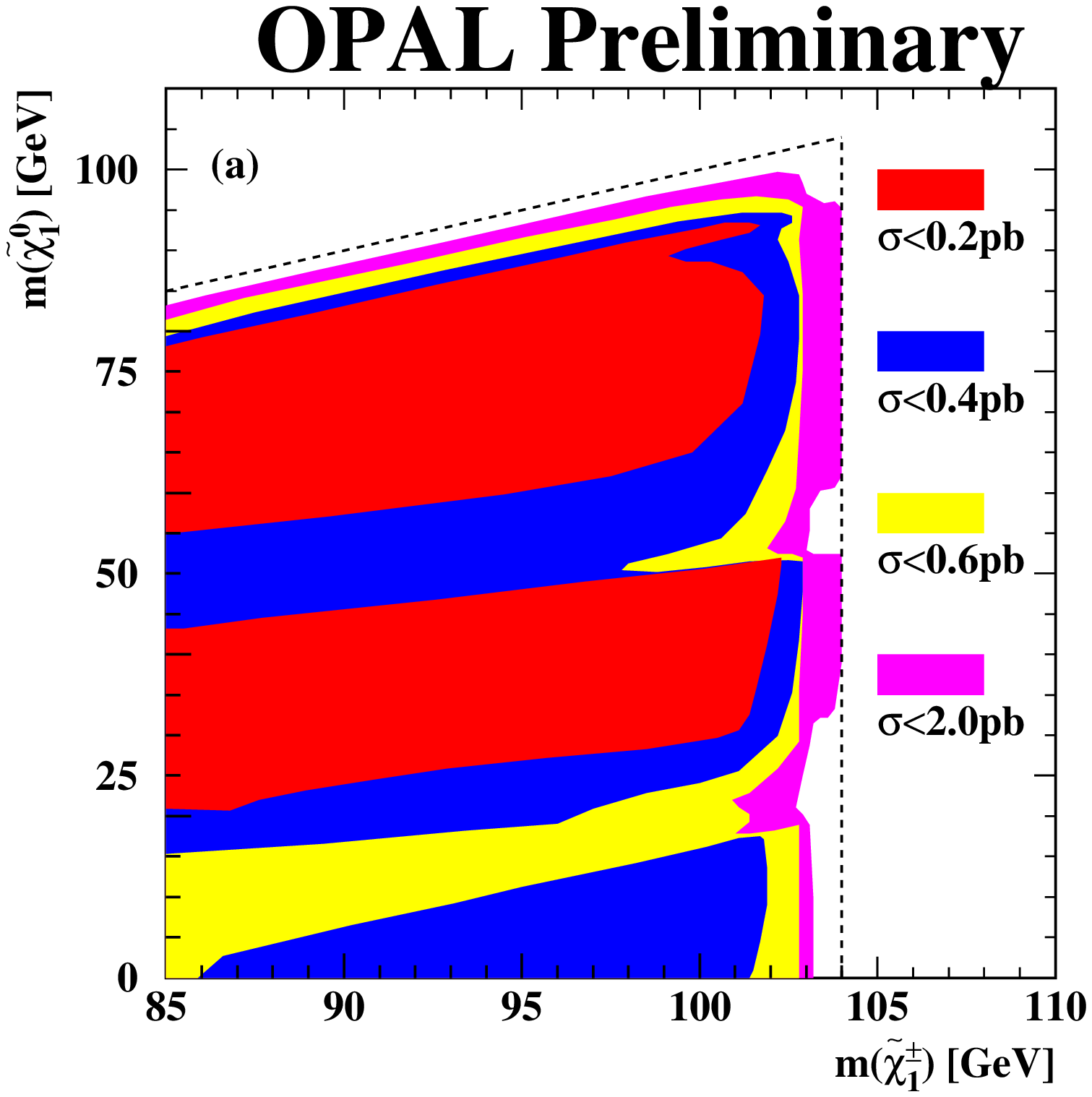,width=0.6\textwidth}}
\vspace{-0.4cm}
\centerline{\epsfig{file=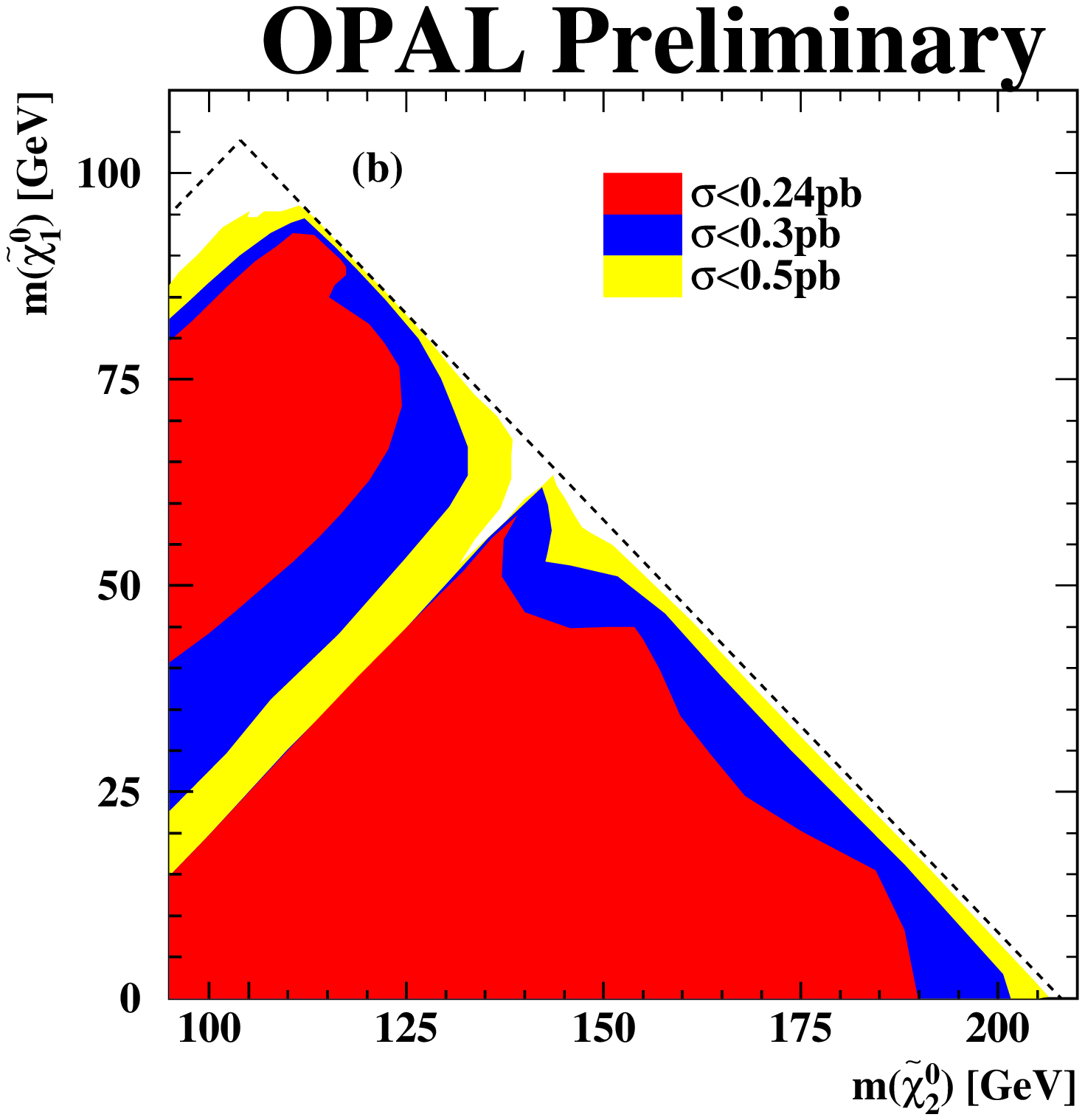,width=0.6\textwidth}}
\caption[]{\label{fig:chargino} 
Contours of the 95\% C.L. upper limits
for (a) the $\ee \ra \chpone \chmone$ production cross-sections
at $\sqrt{s} =$ 208~GeV
are shown
assuming Br$(\chpone \ra \ntone {\mathrm W}^{(*)+}) = 100$\%. 
(b) the $\ee \ra \nttwo \ntone$ production cross-sections at
$\sqrt{s} =$ 208~GeV are shown
assuming Br$(\nttwo \ra \ntone {\mathrm Z}^{(*)0}) = 100$\%. 
The region for which $m_{\nttwo} + m_{\ntone} < m_{\mathrm Z}$
is not considered in this analysis.
The limits use only the data taken in 2000.
}
\end{figure}

An important feature of the chargino searches is that the limits obtained approach
the maximum possible kinematic limits rapidly due to the high expected production
cross-sections, and so the extra data taken at $\sqrt{s}\ge 208$~GeV is very useful
in these searches.  OPAL presents limits on the chargino production cross section
in Figure~\ref{fig:chargino}.  No evidence for a signal is observed, although 
for the search with $\Delta M\approx 10$~GeV, there is an excess observed~\cite{OPALsep5} in the
OPAL experiment: five events are counted in
the data, while 0.74 events are expected from the sum of all Standard Model backgrounds.
None of the other experiments sees a similar excess, and the significance is diluted by
the fact that many different search regions in four experiments were independently investigated
and that a fluctuation can happen in any of them.

Neutralinos may also be produced via $s$-channel \Zo\ exchange, or via $t$-channel
selectron exchange, and the lightest neutralino \co\ is assumed to be the lightest
supersymmetric particle (LSP).  Pair production of \co\co\ is impossible to detect aside from
the signature of the residual initial state radiation.  Instead, associated production of
\ct\co\ is sought, where \ct\ra\co\Zo, and the \Zo\ decay products are observed.
The observable final states are then two jets and missing energy, or two leptons and
missing energy, with similar backgrounds to the chargino searches.  OPAL's limits on
the neutralino production cross-section are shown in Figure~\ref{fig:chargino}.

Searches for sleptons similarly focus on the final state of two like-flavored,
opposite-signed leptons with missing energy, produced by the process
\ee\ra\slsl\ followed by \slp\ra\co$\ell^+$.  In the 1999 data, there was an
excess of events passing the requirements of the stau searches in all four detectors -- no single
experiment had a significant effect, but in combination the signinficance was
greater: $1-{\mathrm{CL}}_b=0.001$ when 1998 and 1999 data were combined~\cite{LEPSUSY1999},
with a stau mass hypothesis of 85~GeV and a neutralino mass hypothesis of 22~GeV.
However, the excess did not persist in the 2000 data collected by the four
experiments, and the particular hypothesis mentioned above is now excluded at the
95\% CL~\cite{LEPSUSYY2K}.

After the July~20 LEPC presentation by ALEPH reporting an excess in 
a preliminary search~\cite{ALEPHjulylepc} for a very light sbottom (of mass between 3 and 4 GeV),
the DELPHI and OPAL collaborations performed similar searches.  OPAL
sees a deficit of events, with 15 events observed and 20.5 events expected from
Standard Model background processes~\cite{karaosaka}.  DELPHI similarly
does not see an excess~\cite{DELPHIsep5}.  The ALEPH experiment updated the search
with an improved Monte Carlo and a lepton identification algorithm which is more appropriate
for identifying leptons inside dense jets and does not report a significant excess, with 24 events
observed and 20 events expected from Standard Model backgrounds~\cite{ref:schlattertalk}.

\section{The Photon Recoil Spectrum}

\begin{figure}[p]
\centerline{\epsfig{file=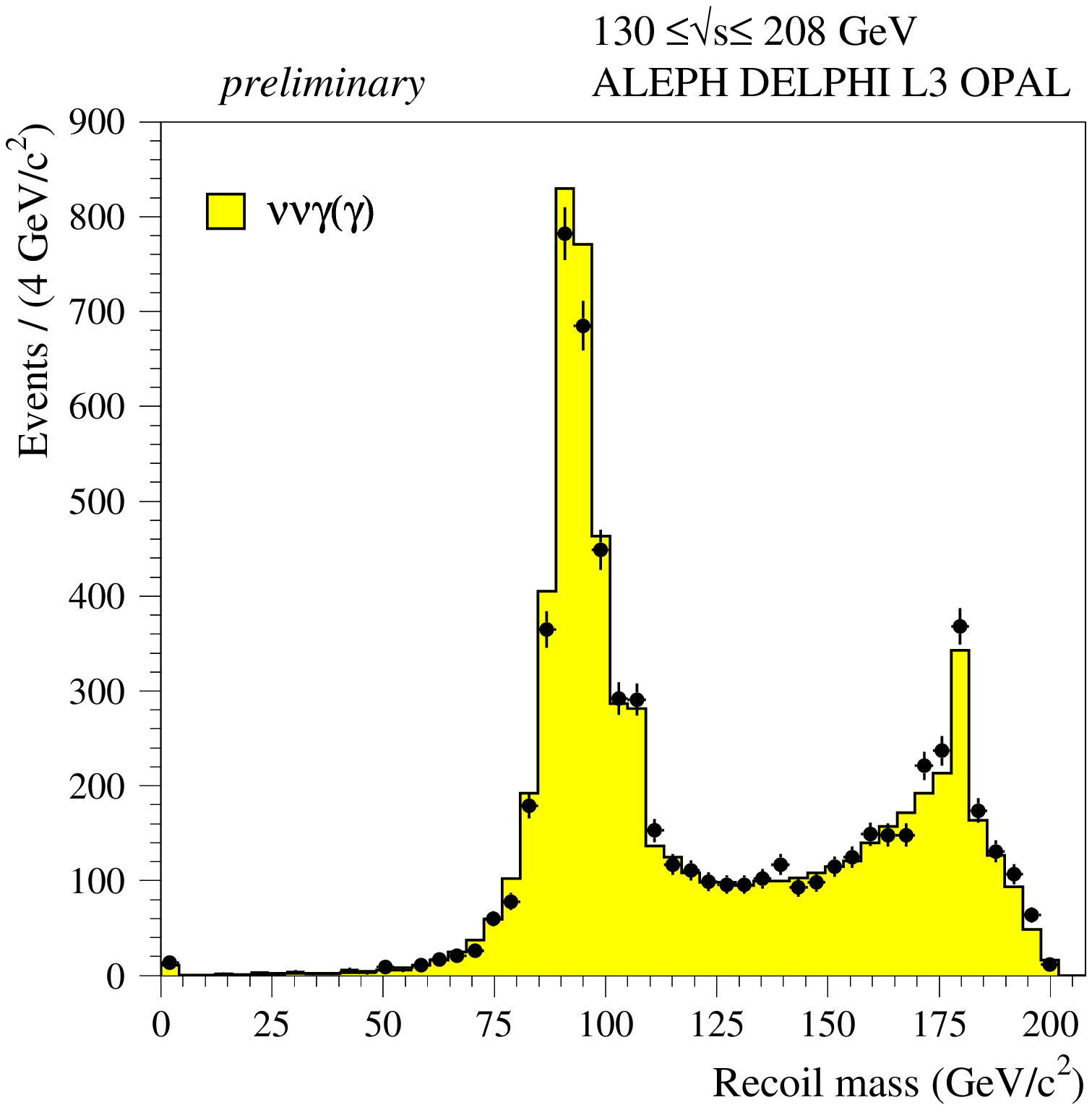,width=0.6\textwidth}}
\centerline{\epsfig{file=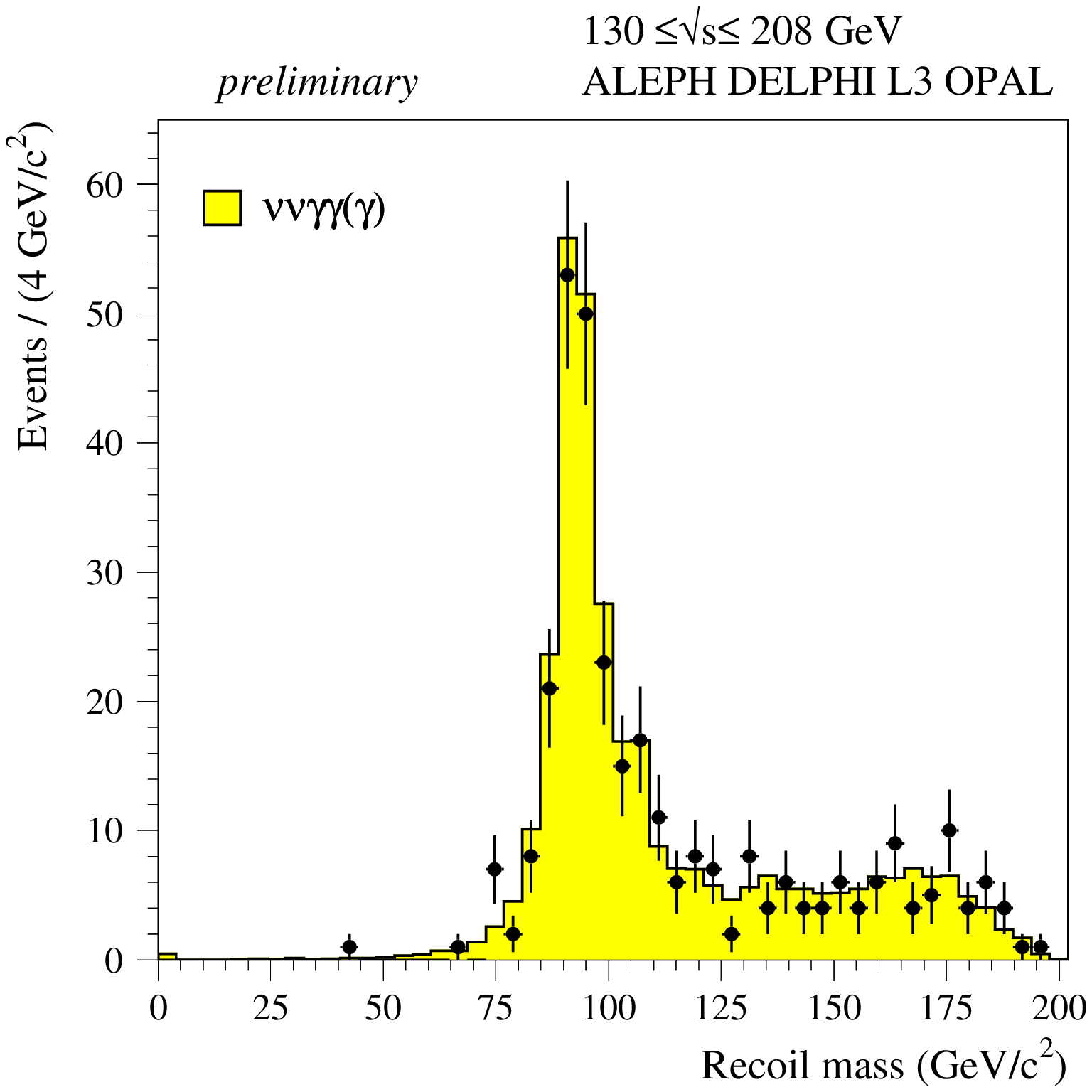,width=0.6\textwidth}}
\caption[]{\label{fig:photonrecoil} 
Spectrum of the recoil mass in events with one single high-energy photon (top)
and with two or more photons (bottom).  Data 
with $130\le\sqrt{s}\le 208$~GeV from the four LEP experiments
are combined.
}
\end{figure}

Events containing a single high-energy photon are valuable for searching for non-interacting
new particles, such as LSP neutralinos (already mentioned), or 
\ct\ra\co$\gamma$.  In Gauge-mediated SUSY-breaking models, the gravitino can be the
LSP, and \co\ra$\tilde G \gamma$ is possible.  An excited neutrino may
decay radiatively.  In general, any invisible process may also be accompanied by initial
state radiation which may be detected, giving a sign for new physics.
Unfortunately, no excess is observed in the recoil mass spectrum to single and multiple
photons, as shown in Figure~\ref{fig:photonrecoil}, which combines~\cite{LEPSUSYY2K} the four
LEP experiments' results for all data taken with $\sqrt{s}\ge130$~GeV.

\section{Prospects for Further LEP Running}

The hint of an excess in the Standard Model Higgs searches near with \mH=115~GeV
has generated a good deal of interest in extending the LEP run through 2001 with
an energy upgrade.  

Since the RADCOR2K conference, there have been two additional updates of the significance
of the SM Higgs search results, at the LEPC presentations of October 10, 2000~\cite{tomlepfest}
and on November 3, 2000~\cite{piktalk}, with significances reported of 2.5$\sigma$ and 2.9$\sigma$,
respectively.  Some variation is expected in both the signal and background cases due to
statistical fluctuations -- large jumps in the significance occur with the discrete arrival
of candidates with large local values of $s/b$.

At the November~3 LEPC open session, the ALEPH, DELPHI, L3 and OPAL collaborations and the LEP
Higgs Working Group jointly recommended running LEP in 2001.  On the same day, the LEP Experiments
Committee met in a closed session and was undecided on the recommendation, balancing the
construction schedule, the cost, and the staffing of the LHC against the LEP run request.  The research board
also failed to make a recommendation, and on November 8, 2000, a press release was issued that LEP
was closed, and on November 15, 2000 a committee of the CERN council was convened, which also failed
to endorse the run extension request.  Dismantling LEP began in early December, 2000.

\section*{Acknowledgements}

The LEP accelerator has performed outstandingly well due to the hard work of
the members of the accelerator divisions at CERN.  In particular, the
RF performance has improved throughout the year, miniramps and the
the bending-field spreading technique were commissioned, and the accelerator
had a high reliability and short turnaround times, all due to hard work
and good organization.  The author would like to thank the members of the four LEP collaborations,
and the members of the LEP Higgs working group for the exemplary cooperation and openness
in supplying search results for combination.

\clearpage
\newpage

\begin{table}
\begin{center}
\begin{tabular}{|l|c|c|c|}
\hline
 Experiment & Data & Background & Signal \\ \hline
ALEPH  & 7 & 3.3 & 1.0 \\
DELPHI & 5 & 5.4 & 1.3 \\
L3 & 4 & 4.0 & 0.3 \\
OPAL & 11 & 9.6 & 0.9 \\\hline
{\bf{LEP}} & {\bf{27}} & {\bf{22.2}} & {\bf{3.6}}\\\hline
\end{tabular}
\end{center}
\caption[]{\label{tab:smexpt}
Numbers of observed events and the expected event counts from a 114~GeV
Higgs boson signal and the Standard Model background processes.}
\end{table}

\begin{table}
\begin{center}
\begin{tabular}{|l|c|c|} \hline
Experiment & Observed (GeV) & Expected (GeV) \\ \hline
ALEPH & 109.1 & 112.5 \\
DELPHI & 110.5 & 110.9 \\
L3 & 108.8 & 110.2 \\
OPAL & 109.5 & 111.7 \\\hline\hline
Channel & Observed (GeV) & Expected (GeV) \\ \hline
Leptons & 109.9 & 108.8 \\
Neutrinos & 112.1 & 110.7 \\
Taus & 105.4 & 104.2 \\
Four Jets & 109.0 & 113.5 \\\hline
{\bf{LEP}} & {\bf{112.3}} & {\bf{114.5}} \\\hline
\end{tabular}
\end{center}
\caption[]{\label{tab:smlimit}
Limits on the mass of the Higgs boson, assuming the Standard Model 
production cross-section and branching fractions, by experiment, by channel, 
and combined.  These have been computed with a uniform procedure and
may vary by small amounts from the ones quoted by the individual experiments.  
In the lepton channel, there is a small unexcluded region below 100.7 GeV.}
\end{table}

\begin{table}
\begin{center}
\begin{scriptsize}
\begin{tabular}{|c|c|c|c|c|c|c|}\hline
Scenario & \mh\ limit & \mh\ limit & \mA\ limit & \mA\ limit & $\tan\beta$ limit & 
$\tan\beta$ limit \\
 & obs (GeV) & exp (GeV) & obs (GeV) & exp (GeV) & obs & exp \\\hline
\mh-max & 89.5 & 93.8 & 90.2 & 94.1 & 0.53--2.25 & 0.48--2.48 \\\hline
No stop mix & 89.4 & 94.3 & 89.6 & 94.6 & 0.9--7.2 & 0.8--15 \\\hline
\end{tabular}
\end{scriptsize}
\end{center}
\caption[]{\label{tab:mhmalimits}
Limits on \mh, \mA, and $\tan\beta$ in the \mh-max and no-mixing scenarios.  
The limits obtained by the combination of the four LEP experiments' data are
indicated with ``obs," while the median limits expected to be obtained in a large
ensemble of background-only experiments are labeled ``exp."}
\end{table}

\clearpage
\newpage

\end{document}